\documentclass[fleqn,10pt]{wlscirep}
\usepackage[utf8]{inputenc}
\usepackage[T1]{fontenc}
\usepackage[normalem]{ulem}
\usepackage{multirow}
\usepackage{lineno}

\newcommand{\red}{\textcolor{red}}

\newcommand{\RomanNumeralCaps}[1]
{\MakeUppercase{\romannumeral #1}}

\title{Polariton lasing in Mie-resonant perovskite nanocavity}

\author[1,2,+]{M.A.~Masharin}
\author[2,+]{D.~Khmelevskaia}
\author[2]{V.I.~Kondratiev}
\author[2]{D.I.~Markina}
\author[2]{A.D.~Utyushev}
\author[2]{D.M. Dolgintsev}
\author[2]{A.D.~Dmitriev}
\author[2]{V.A.~Shahnazaryan}
\author[2]{A.P.~Pushkarev}
\author[1,4]{F.~Isik}
\author[2]{I.V.~Iorsh}
\author[2,3]{I.A.~Shelykh}
\author[1,4*]{H.V.~Demir}
\author[2,5*]{A.K.~Samusev}
\author[2,6*]{S.V.~Makarov}

\affil[1]{UNAM-Institute of Materials Science and Nanotechnology, National Nanotechnology Research Center, Department of Electrical and Electronics Engineering, Department of Physics, Bilkent University, Ankara, 06800, Turkey}
\affil[2]{ITMO University, School of Physics and Engineering, St. Petersburg, 197101, Russia}
\affil[3]{Science Institute, University of Iceland, Dunhagi 3, IS-107, Reykjavik, Iceland}
\affil[4]{LUMINOUS! Center of Excellence for Semiconductor Lighting and Displays, School of Electrical and Electronic Engineering, School of Physical and Mathematical Sciences, School of Materials Science and Engineering, Nanyang Technological University, 639798, Singapore}
\affil[5]{Experimentelle Physik 2, Technische Universit\"at Dortmund, 44227 Dortmund, Germany}
\affil[6]{Qingdao Innovation and Development Center, Harbin Engineering University, Qingdao 266000, Shandong, China}

\affil[+]{Equal contribution}
\affil[*]{Corresponding authors}

\begin{abstract}
    
Deeply subwavelength lasers (or nanolasers) are highly demanded for compact on-chip bioimaging and sensing at the nanoscale. One of the main obstacles for the development of single-particle nanolasers with all three dimensions shorter than the emitting wavelength in the visible range is the high lasing thresholds and the resulting overheating. Here we exploit exciton-polariton condensation and mirror-image Mie modes in a cuboid CsPbBr$_3$ nanoparticle to achieve coherent emission at the visible wavelength of around 0.53~$\mu $m from its ultra-small ($\approx$0.007$\mu$m$^3$ or $\approx\lambda^3$/20) semiconductor nanocavity. The polaritonic nature of the emission from the nanocavity localized in all three dimensions is proven by direct comparison with corresponding one-dimensional and two-dimensional waveguiding systems with similar material parameters.
Such a deeply subwavelength nanolaser is enabled not only by the high values for exciton binding energy ($\approx$35~meV), refractive index ($>$2.5 at low temperature), and luminescence quantum yield of CsPbBr$_3$, but also by the optimization of polaritons condensation on the Mie resonances. Moreover, the key parameters for optimal lasing conditions are intermode free spectral range and phonons spectrum in CsPbBr$_3$, which govern polaritons condensation path. Such chemically synthesized colloidal CsPbBr$_3$ nanolasers can be easily deposited on arbitrary surfaces, which makes them a versatile tool for integration with various on-chip systems.

\end{abstract}

\begin{document}

\flushbottom
\maketitle
\section*{Introduction}

Nowadays, compact lasers are gradually replacing or complementing electrical interconnects, with billions of devices deployed since 1990-s~\cite{tatum2015vcsel}. In turn, the increase of optical elements on a photonic chip is another technological trend requiring integrated lasers with the smallest possible volume~\cite{smit2012moore, agrell2016roadmap}. Thus, continuing laser sources miniaturization is a natural approach to lowering energy consumption and creating compact devices with various functionalities~\cite{ma2014explosives, ma2019applications, liang2020plasmonic}.  
Pioneering works on near- and subwavelength lasers date back to the turn of the century with the demonstration of microdisk lasers~\cite{mccall1992whispering}, photonic crystal lasers~\cite{painter1999two} and nanowire lasers~\cite{johnson2001single}. In these and many other designs~\cite{hill2014advances,kodigala2017lasing,ha2018directional, mylnikov2020lasing, azzam2020ten, wu2021bound, tripathi2021lasing, hwang2021ultralow, sung2022room}, optical modes are confined in one or two dimensions, where one of the sizes remains larger than the mode wavelength. The nanolasers where all geometrical parameters are truly subwavelength were demonstrated in a very limited number of works due to high technological requirements to materials and cavity design~\cite{nezhad2010room,khajavikhan2012thresholdless,ding2012metallic}. Indeed, these limitations prevent the creation of all-subwavelength nanolasers for visible range, except colloidal nanolasers where observation of lasing from a single nanolaser is still a challenge. In turn, a polariton laser does not require inversion of population and, thus, can work at much lower charge carrier densities as compared with classical lasers~\cite{byrnes2014exciton}. 

One of the most suitable light-emitting materials for supporting a polariton lasing is cesium lead bromide CsPbBr$_3$ perovskite owing to high photoluminescence (PL) quantum yield up to 95\% \cite{di2017near}, defect tolerance~\cite{kang2017high}, strong excitons~\cite{protesescu2015nanocrystals,baranowski2020excitons} with a high refractive index around the excitonic state~\cite{ermolaev2022giant}, high gain~\cite{tatarinov2023high}, and ultrafast optical response~\cite{huang2020ultrafast}. Polariton condensation and lasing in various CsPbBr$_3$ cavities was shown in CsPbBr$_3$ nanowires~\cite{evans2018continuous}, microplates~\cite{tao2022halide}, and more complex systems~\cite{su2020observation, su2021perovskite, wu2021nonlinear,feng2021all, li2022manipulating}. Remarkably, phonons in CsPbBr$_3$ containing heavy ions possess energies just up to 38~meV,~\cite{zhao2019simultaneous} which can play a crucial role in the polaritons condensation in the case of small cavities with large inter-mode spectral separation, paving the way to the efficient operation of nanolasers.

In this work, we exploit an inversion-free regime for a polariton nanolaser with a record small volume $(\lambda/2.7)^3$ among all known perovskite lasers. Here the polariton nature of lasing is proved experimentally based on the direct comparison of 2D, 1D, and 0D perovskite waveguiding and resonant systems, as well as by analytical modeling. The detailed experimental study and numerical full-wave calculations allow us to find a cavity design made of CsPbBr$_3$ with the smallest volume supporting lasing at low temperature. Namely, a mirror-image quadrupole Mie mode in a nanocuboidal CsPbBr$_3$ particle on silver film allows for a high-enough quality factor ($Q$) and inter-mode spectral distance, which is larger than the highest energy of optical phonon in CsPbBr$_3$. From a technological point of view, the developed method for colloidal synthesis of the lasing nanoparticles (NPs) enables the simple deposition of nanolasers, which makes it suitable for on-chip integration with various photonic circuits.

\section{Results}

\subsection{Concept of polariton stimulated relaxation in different perovskite structures.}

\begin{figure}[t!]
\centering
\center{\includegraphics[width=1\linewidth]{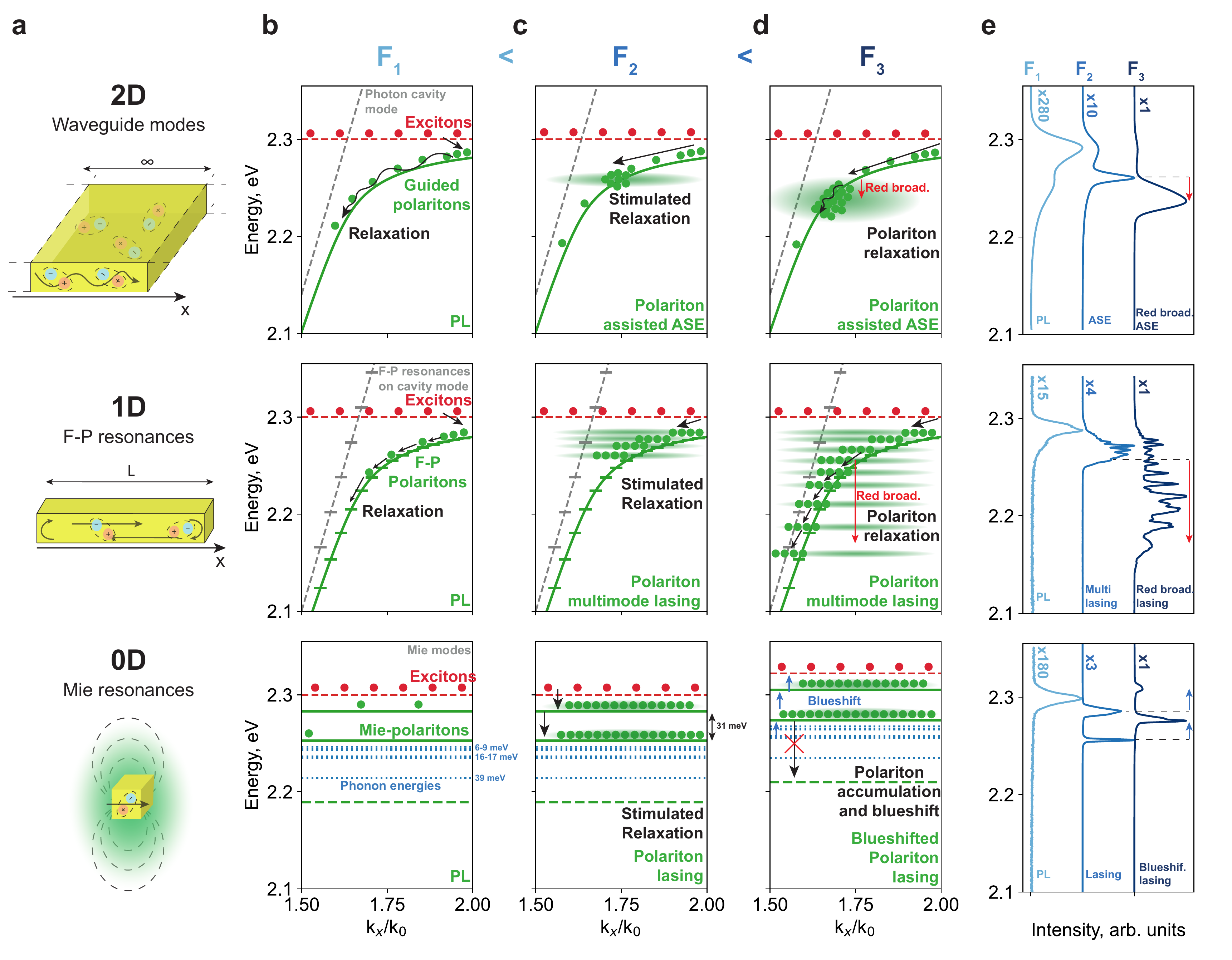}}
\caption{\textbf{The concept of the polariton stimulated emission in low-dimensional perovskite photon cavities.} (a) Illustration of studied perovskite structures: thin film (2D), nanowire (1D) and nanocube (0D). 
(b) Scheme of the PL linear regime, estimated photon cavity modes (grey dashed lines) strongly coupled with exciton resonance (red dashed lines) resulting in exciton-polariton formation (green lines) for each of the structures. Waveguide mode in 2D perovskite film results in guided polariton; Quantized Fabry-Perot (F-P) resonances appeared from waveguide mode in 1D nanowires resulting in F-P polariton resonances; Mie resonances supported in 0D nanocubes results in Mie-polaritons. Red circles are excitons in the systems, and green circles are exciton-polaritons. 
(c) Scheme of the stimulated polariton relaxation appeared in the different systems with increasing the pump fluence F$_2$ > F$_1$ which leads to the appearance of ASE in 2D thin films, multimode lasing in 1D nanowire structure, and few-mode lasing in 0D nanocubes. 
(d) Scheme of the polariton interaction at higher pump fluence F$_3$>F$_2$. Polaritons in 2D and 1D  relax at lower energy, and in 0D they accumulate and shift the level to the blue region. (e) Experimentally measured normalized emission spectra at pump fluences F$_1$, F$_2$, F$_3$ from each of the perovskite structures. The colors of the spectra correspond to the fluences, imaged in (b-d).}
\label{fig1}
\end{figure}

First, let us discuss the concept of polariton stimulation in a nonlinear regime in perovskite thin films, nanowires, and nanocubes (Fig.~\ref{fig1}a) upon a non-resonant (i.e. excited by photons with the energy higher than that for the exciton state) optical pump. Each of these perovskite structures supports exciton-polariton states, schematically shown in Fig.~\ref{fig1}b, where a guided polariton in the film has a dispersion curve below the light cone \cite{sich2012observation}; in a nanowire, the waveguided mode has quantized F-P resonances \cite{shang2020role}; and polaritons in a Mie-resonant cavity do not have angular dispersion, but red-shifted from their uncoupled states \cite{wang2020collective}.
Exciton-polaritons in these structures can be excited non-resonantly by a femtosecond (fs) laser pulse, as we show experimentally in the next sections. Varying the pump fluence one can control the number of polaritons in the system. At the low pump fluences around $F_1$, the photoexcited carriers thermalize forming excitons and exciton-polaritons in the available states (red and green dots, respectively, in Fig.~\ref{fig1}b). The formed polaritons relax to lower energies through phonon scattering and then recombine \cite{byrnes2014exciton}. Nonlinear properties start to play a significant role at the higher pump fluences $F_2$ when stimulated relaxation and hence stimulated emission appears. Large exciton and polariton concentrations cause nonlinear boson stimulation~\cite{deng2010exciton}, which leads to the polariton accumulation at some specific state on the polariton branch, where polaritons still have high exciton fraction, but low non-radiative losses (Fig.~\ref{fig1}c). The recombination of the accumulated polaritons, which occupy one quantum state, provides the coherence of the emission. At this moment one can mention there are no observable differences between the stimulated polariton emission and the well-known amplified spontaneous emission (ASE) except for the origin of the coherence \cite{byrnes2014exciton}. However, for even higher pump fluences $F_3$, polaritons start interacting at the accumulated levels, leading to different phenomena. If polariton states at lower energies exist, like for guided polaritons or F-P polaritons, the polaritons scatter to the lower energy states \cite{shan2022brightening}. In the case of perovskite nanocubes, where there can be no states or the difference between the neighboring levels is larger than existing phonon energies, polaritons are not able to thermalize. Therefore, they stronger accumulate in larger numbers at the available states, which leads to the polariton-polariton Coulomb interaction and polariton blueshift, observed in polariton systems (Fig.~\ref{fig1}d) \cite{su2020observation,su2021perovskite}. 

The discussion of the concept can be supported by the bosonic kinetic model, limiting to the three discrete polariton levels for simplicity. We attribute each of the levels to a quantized excitonic eigenmode. The non-resonant pulsed excitation creates a large non-equilibrium population of electrons and holes, forming the exciton reservoir with the number of excitons $n_R$, which plays as a feeding source for the three bright exciton levels with energies $E_{X1} < E_{X2} < E_{X3}$. We also consider the transitions between different excitonic levels. As the energy gaps between the peaks are of the order of tens of meV, we assume that optical phonons mediate these transitions.
In all the transitions, we take into account the bosonic nature of excitons, resulting in $\propto n_i (n_j+1)$ type terms.
The system of equations describing the dynamics then reads 

\begin{align}
    \frac{{\rm d} n_R }{{\rm d}t} 
    &= P(t) - \left[ \gamma_R + \alpha_3 (n_3+1) + \alpha_2 (n_2+1) + \alpha_1 (n_1+1) \right] n_R  \notag \\
    \frac{{\rm d} n_3 }{{\rm d}t} 
    &= \alpha_3 n_R (n_3+1)
    -\left( \frac{1}{\tau_3^r} + \frac{1}{\tau_3^{nr}} +\gamma^{(2)} n_3 \right) n_3
    \notag \\
    &+\frac{1}{\tau_{32}} \left[ e^{-\beta \Delta_{32}} n_2 (n_3+1) - n_3 (n_2+1) \right]
    +\frac{1}{\tau_{31}}  \left[ e^{-\beta \Delta_{31}} n_1 (n_3+1) - n_3 (n_1+1) \right] 
    \notag \\
    \frac{{\rm d} n_2 }{{\rm d}t} 
    &= \alpha_2 n_R (n_2+1)
    -\left( \frac{1}{\tau_2^r} + \frac{1}{\tau_2^{nr}} +\gamma^{(2)} n_2 \right) n_2
    \notag \\
    &+\frac{1}{\tau_{32}}  \left[n_3 (n_2+1) - e^{-\beta \Delta_{32}} n_2 (n_3+1)   \right]
    +\frac{1}{\tau_{21}}  \left[ e^{-\beta \Delta_{21}} n_1 (n_2+1) - n_2 (n_1+1) \right]
    \notag \\
    \frac{{\rm d} n_1 }{{\rm d}t} 
    &= \alpha_1 n_R (n_1+1)
    -\left( \frac{1}{\tau_1^r} + \frac{1}{\tau_1^{nr}} +\gamma^{(2)} n_1 \right) n_1 \notag \\
    &+\frac{1}{\tau_{31}}  \left[n_3 (n_1+1) - e^{-\beta \Delta_{31}} n_1 (n_3+1)   \right]
    +\frac{1}{\tau_{21}}  \left[n_2 (n_1+1) - e^{-\beta \Delta_{21}} n_1 (n_2+1)  \right]
    \label{eq_model}
\end{align}
where $P(t) = P_0 e^{-(t-t_0)^2/(2\tau_p^2)}$ stands for pulsed non-resonant pump, $\alpha_i$ denote the scattering rate from the reservoir to the $i$-th excitonic level, $\tau_i^{r[nr]}$ are radiative [non-radiative] lifetimes of excitons, and $\gamma^{(2)}$ is the exciton-exciton annihilation  rate. 
Here $\beta =1/k_B T$, with $k_B$ being the Boltzmann constant, $\tau_{ij}$, $\Delta_{ij}=|E_i - E_j|$ denote the transition rates and the energy gaps between different exciton levels.
The respective emission is estimated as $I_i \propto \int n_i(t) {\rm d} t / \tau_i^r$. 
The results of the simulation will be demonstrated in the next section and here we qualitatively discuss the process.

At the low amplitude of $P(t)$, only linear terms of $n_R$ play a significant role as a source of the particles for all three levels, and linear terms of radiative and non-radiative lifetimes $\tau_{1-3}^{r,nr}$ provide the quasiparticles annihilation. After some threshold the nonlinear term of $n_R n_3$ in the second equation of the system Eq.(\ref{eq_model}) causes stimulated relaxation of the quasiparticles to the $n_3$ level, resulting in polariton stimulated emission. At the same time, it dramatically increases the rate of the quasiparticle scattering from $n_3$ to $n_2$ with a characteristic time of $\tau_{32}$. With further increase of the $P(t)$ and, therefore $n_R$, the terms of $1/\tau_{32} n_3 n_2$ and $\alpha_2 n_R n_2$ cause the stimulation to the lower $n_2$, providing the stimulated emission from the second level. Similarly, at some higher pump threshold, the stimulated relaxation reaches the $n_1$ via quadratic $n$ terms in the equations. The values of the thresholds are determined by the characteristic scattering times $\tau$ and the rate of the scattering from the reservoir $n_R$ to the polariton levels $\alpha$, which depend on the phonon energies and the exciton fraction in the polariton. This approach of the three-level model is implemented for the 0D case of nanocuboid lasing, well corresponding to the experimental observations (dashed lines in Figure \ref{fig3}h). Moreover, in 1D and 2D systems, the process of the stimulated polariton relaxation can be described in a similar way, but with a continuous spectrum of the polariton states.

In order to qualitatively support the above-mentioned concept of polariton lasing, we demonstrate measured emission spectra integrated over angles and obtained under non-resonant fs pump with different pump fluences for lead-bromide perovskite thin film, nanowire, and nanocuboid (Fig.~\ref{fig2}a, \ref{fig2}f and \ref{fig2}k, respectively) at $T=6$~K in Fig.\ref{fig1}e. At pump fluence $F_1$ in the linear regime, we observe linear PL spectra with tails in the red region. However, it should be noted that the asymmetric PL spectra origin is still under discussion and could be originated from other effects, such as defect states and PL re-absorption.\cite{chen2018excitonic,saba2016excited,ryu2021role}. Increasing pump fluence up to $F_2$ ASE in the film, multimode lasing in the nanowire, and few-mode lasing in nanocuboid appear. According to the concept discussed above stimulated emission shifts to the red region in 2D and 1D structures and to the blue region in 0D at the highest $F_3$ pump fluence (Fig.~\ref{fig1}e). 

To experimentally check if we are below the Mott transition, we further increase the pump fluence up to 3 mJ/cm$^2$ for the nanocuboids and observe the broad PL spectrum without lasing modes with the shifted central peak from the initial exciton PL up to the 32 meV, which is very close to the value of exciton binding energy (see Fig.~S4 in SI for the details). Nevertheless, even under the Mott transition electron-hole pairs can still be correlated to Bardeen–Cooper–Schrieffer (BCS) states, which strongly couples to the optical mode, resulting in BCS polaritons.\cite{enomoto2022drastic} This phenomenon allows observing polariton stimulation and condensation even above the Mott transition. Experimentally collected data, shown in Fig.\ref{fig1}e, supports the polariton stimulation model only indirectly and in the following sections, we show more experimental insights, supporting the idea.


According to the conventional approach, the spectral position of the optical gain profile and ASE shifts to the blue region and grows in intensity with increasing pump fluence in semiconductors. \cite{klimov2000optical,hofmann2002emission} Nevertheless, in perovskite structures, we observe controversial results (Fig.~\ref{fig1}e), which can be explained by exciton-polariton relaxation discussed above. In order to check the assumption we have to get rid of exciton-polaritons in the system with lead-bromide perovskites. As the exciton state in the material is stable at room temperature and the high refractive index provides strong field localization in a variety of bulk perovskite structures, we can turn to the colloidal perovskite nanoparticles. In this case, random fluctuations of the nanoparticles in the solution prevent strong light-matter coupling and can be considered a non-polariton system. Indeed, the measured gain profile in the system in Ref.~\cite{geiregat2018using} represents a clear conventional gain profile blueshift with pump fluence increasing rather than red broadening.

The role of exciton-polaritons in the lead-bromide perovskite structures is important from a fundamental understanding of the process and it also reveals new efficient approaches for the nanolaser designs. In the following sections, we properly prove the presence of exciton-polaritons in the studied structures and suggest a new approach for efficient inversion-free nanolasers based on the Mie-polaritons.

\subsection{Exciton-polariton stimulated emission in different perovskite structures.}

\begin{figure}[t!]
\centering
\center{\includegraphics[width=0.9\linewidth]{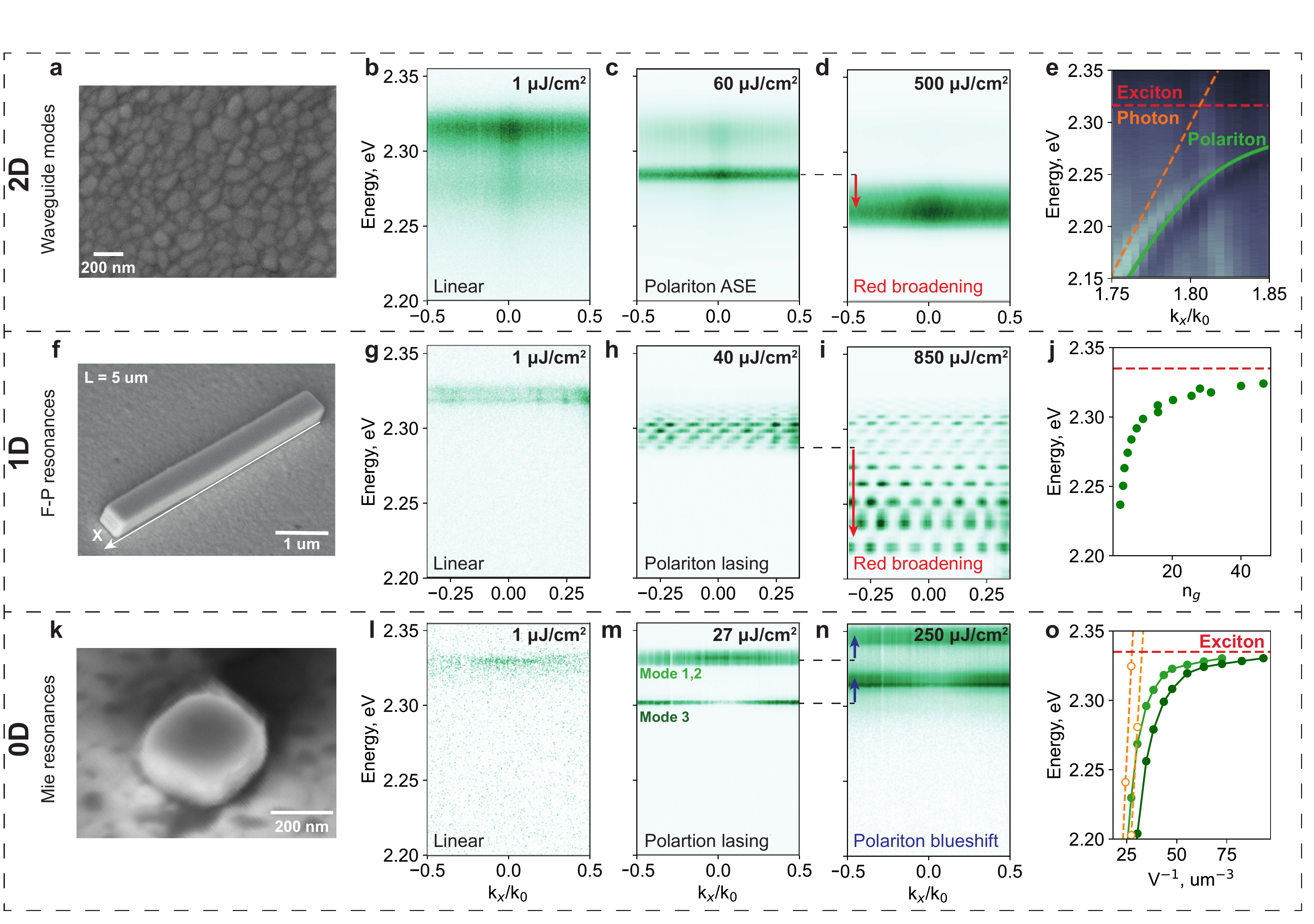}}
\caption{\textbf{Experimental results of the stimulated emission of perovskite thin film, nanowire, and nanocuboid.} (a) SEM image of the studied perovskite thin film. (b-d) Angle-resolved emission spectra of the thin film, obtained under a non-resonant fs pump at 6~K for pump fluences of 1, 60, and 500 $\mu$J/cm$^2$, which correspond to the linear PL, polariton ASE, and red broadened polariton ASE, respectively. (e) Angle-resolved reflection spectra of the polariton-guided mode, measured at room temperature below the light cone. The dashed orange line shows the estimated uncoupled photon mode, the dashed red line shows the estimated exciton level, and the solid green line shows the fitted polariton mode. (f) SEM image of the studied perovskite nanowhisker with a length of around 5 $\mu$m. (g-j) Angle-resolved emission measurements of the nanowire for pump fluences of 1, 40, and 850 $\mu$J/cm$^2$, respectively. The data shows the dynamics of polariton multimode lasing emission. (j) The group refractive index of the studied nanowire as a function of the energy, estimated from the FSR of lasing peaks (See SI for the details).
(k) SEM image of the studied perovskite nanocuboid. (l-n) Angle-resolved emission measurements of the nanocuboid for pump fluences of 1, 27, and 250 $\mu$J/cm$^2$, respectively. (o) The estimated mode spectral center as a function of the inverted volume.
}
\label{fig2}
\end{figure}

First, we experimentally study exciton-polaritons in 2D perovskite structures - polycrystalline thin films, which are synthesized by solvent engineering method (See Methods for the details). Synthesized films, shown in Fig.~\ref{fig2}a, have roughness equal to 5.1 nm, for 120 nm film width (See Fig. S12 in SI for more details).

We use non-resonant fs emission with a photon energy of 2.53 eV and 100 kHz of repetition rate to measure emission spectra with angular resolution (Figs.~\ref{fig2}b-c). At low pump fluence around 1 $\mu$J/cm$^2$ linear excitonic PL spectrum without any pronounced radiation pattern is observed (Fig.~\ref{fig2}b). The low intensive broad peak in the red region can be attributed to guided polaritons scattered on morphology defects and grain boundaries. With increasing the pump fluence up to 60 $\mu$J/cm$^2$ a wide ASE peak appears at the energy around 2.27 eV. The ASE emission strongly broadens toward the red region with increasing pump fluence (Fig.~\ref{fig2}d), as was discussed in the previous section. 

To confirm the polaritonic nature of the ASE, we directly measure waveguide mode dispersion at room temperature by the angle-resolved spectroscopy method combined with a solid immersion lens (SIL, see Methods for the details). The method allows for measurement of guided mode as a function of the wavenumber below the light cone through leakage tails, collected by the SIL. The angle-resolved reflection spectrum of the thin film below the light cone is shown in Fig.~\ref{fig2}e. In the measured spectra, one can mention pronounced curvature near exciton resonance, which is a sign of the strong light-matter coupling regime. We extract guided mode dispersion from the experimental data and fit it by the two-coupled oscillator model to evaluate a light-matter coupling regime~\cite{hopfield1958theory} (see Methods for the fitting details). As a result, the estimated light-matter coupling coefficient $g_0 = 72~$meV and Rabi splitting $\Omega_{Rabi} = 143~$meV are much larger, than the photon and exciton linewidths, which are estimated to be 30 and 20~meV, respectively. Therefore, the strong coupling regime and guided polariton in the perovskite thin film are confirmed.~\cite{hopfield1958theory} It should be mentioned that the angle-resolved spectroscopy with SIL is possible to realize only at room temperature and the strong light-matter coupling regime is confirmed at these conditions. Moreover, we also observe ASE with angular resolution below the light cone originating from the guided mode and can be controlled by the optical losses (see Section S1 of SI). However, at cryogenic temperatures, non-radiative losses are reduced and exciton fraction in the photoinduced carriers is increasing.~\cite{khmelevskaia2021excitonic} Therefore, if exciton-polaritons exist at room temperature, they have to exist at cryogenic temperatures as well, where all our measurements are carried out.~\cite{masharin2022room}


In recent works, lead-bromide nanowires demonstrated high-quality lasing emission at the room-temperature under pulsed and even CW-pump optical pump.~\cite{pushkarev2018few,shang2020role} Nevertheless, our experiment is conducted at $T=6$~K to study polariton effects with decreased losses properly. According to our assumptions, the nanowire can be considered as a waveguide with quantized F-P resonances and also have to support exciton-polaritons. In this case, the observed ASE should turn to the lasing regime based on the F-P resonances.
We synthesized lead-bromide perovskite nanowires (Fig.~\ref{fig2}f) with a length of around 5~$\mu$m by the method mentioned in the Methods. 
At $T=6$~K in a similar way to the previous experiment, we measure the emission from the nanowire under a femtosecond pump with angle resolution along the axis of the nanowire axis (Figs.~\ref{fig2}g-i). 
At the low pump fluence, we observe only the spontaneous PL emission with uniform intensity distribution over the angle (Fig.~\ref{fig2}g). 
With the increase of pump fluence after the threshold multimode lasing regime is achieved (Fig.\ref{fig2}h). 
The lasing emission has the checkerboard pattern in the angle dependence spectrum, which appears due to the interference of the coherent lasing emission, coming from two edges of the nanowire, and it is discussed in detail in SI. 
With further increase of pump fluence, we observe strong red broadening of the lasing modes up to 100~meV (Fig \ref{fig2}i), the same as for the 2D thin film guided-polariton ASE. 
The origin of the multimode lasing red broadening corresponds to the polariton relaxation and is discussed in the previous section. 

Since there is no option to measure polariton dispersion in a perovskite nanowire properly and prove their existence experimentally, we study the free spectral range (FSR) of lasing peaks, obtained at pump fluence 850~$\mu$J/cm$^2$. 
Thanks to the angle-resolved spectroscopy and checkerboard pattern we can better distinguish one mode from another. 
Based on the FSR we estimate the group refractive index that is rapidly increased up to the value of 45 near the exciton resonance (Fig \ref{fig2}h), which points to the exciton-polariton origin of the lasing F-P resonances, observed before in Ref.\cite{shang2020role} 

To obtain the 0D perovskite system with only a few Mie-polariton states, we prepare lead-bromide nanocuboids via modified hot-injection synthesis and further deposited on the metal-dielectric substrate by drop-casting (See Methods and Section S6 of SI for the details). The substrate with plasmonic effects is used for the enhancement of the field localization and mode selectivity (See Fig. S10 and Fig. S8 in SI), discussed in detail in the next section. 
We perform the same emission measurements for the synthesized perovskite nanocuboid (Fig.~\ref{fig2}k) with a physical volume of 0.02~$\mu m^3$ on the Al$_2$O$_3$/Ag/Si substrate, which supports few Mie-resonances in the spectral region near the exciton level. 
At the pump with fluences lower than 1~$\mu$J/cm$^2$ the low-intensity linear PL spectrum is observed (Fig.~\ref{fig2}l). 
At the pump threshold around 4~$\mu$J/cm$^2$, narrow lasing peaks appear, corresponding to Modes 1 and 2 in Fig.~\ref{fig2}m. With increasing of pump fluence around 17 $\mu$J/cm$^2$ the next Mode 3 appears with pronounced radiation patterns in the angle-resolved spectrum (Fig.~\ref{fig2}m), which corresponds to the particular Mie-modes, discussed in SI, Fig. S3, and also in the next section.

The main difference between the 0D perovskite structure with 1D and 2D is in the dynamics of lasing peaks with increasing pump fluence. 
Instead of stimulated emission red broadening, we observe the blueshift and broadening of the lasing peaks (Fig.~\ref{fig2}n). 
Exciton-polaritons cannot scatter to the lower energy states and therefore accumulates in the available lasing levels. 
Accumulated polaritons in the same quantum state start to interact with each other through Coulomb potential, which shifts polariton levels to the blue region \cite{bajoni2008polariton,gao2012polariton}. 

The broadening of the lasing spectra in nanocuboids is suggested to be originated from two mechanisms. 
The first one is the aforementioned polariton-polariton interaction. 
The second can be caused by temporal broadening: pumping pulse excites a huge number of the polaritons, accumulated in the lasing state; polariton level shifts to the blue region because of the Coulomb interaction; accumulated polaritons fall to the ground state and produce narrow coherent polariton lasing emission; with the time the number of the polaritons in the lasing levels are reducing polariton lasing peak shifts to the red region but with continuing lasing emission. 
The recent work demonstrated the polariton lasing with the angular and temporal resolution, where this phenomenon is observed and confirmed \cite{ardizzone2022polariton}.
As in the experiment, the measurements are performed with time integration, we observe only blueshift and broadening, however, previous studies have already observed the effect in the lead-bromide microlasers with time resolution\cite{schlaus2019lasing}.

Mie-resonances in perovskite nanocuboids strongly depend on the geometrical sizes and have no angular dispersion with the energy of thin film or nanowire. Therefore, we can only indirectly confirm the polariton origin of the Mie-modes by the calculation of eigenfrequencies of Mie-resonances with the account of the refractive index dispersion, affected by exciton resonance. The refractive index at cryogenic temperature is estimated based on the experimental micro-ellipsometry measurements~\cite{ermolaev2022giant} and reflectance spectra of lead-bromide thin nanoplates with different thicknesses measured at room temperature and 6~K (See Fig. S9 in SI for the details). We calculate the spectral position of the Mie-resonance as a function of the nanocuboid volume and compare it with the case when the refractive index is constant for all light wavelengths (See Methods for the details). As shown in Fig.~\ref{fig2}o, with reducing the active media volume the spectral position of the Mie-modes shifts toward the exciton asymptote. If we consider a constant refractive index without exciton resonance, the spectral position of the Mie-modes demonstrates almost linear dependence with the nanocuboid inverted volume. The difference between the two lines can be explained by the strong light-matter coupling regime of the Mie resonances, calculated with the material dispersion and exciton state. The modeling results together with the experimental study of lasing peaks evolution confirm the polariton origin of Mie-modes. 

On one hand, the exciton-polariton nature of perovskite self-resonating cavities allows realizing inversion-free coherent emission due to the boson stimulation. On the other hand, polariton relaxation leads to the red broadening of coherent emission. The latter does not allow the achievement of a single or few-mode lasing regime and also prevents effective polariton accumulation in a single state with narrow lasing emission. However, if we turn to the 0D structure, where there are no continuous polariton states and only discrete Mie resonances, the phenomenon of the stimulated emission red broadening  disappears, while the inversion-free regime would remain. In the next sections, we represent the clear transition from the multi-mode lasing regime to the single-mode lasing in perovskite nanocuboid, show the analysis of the Mie-modes in perovskite nanocuboids, located on the plasmonic substrate, and study in detail the dynamics of the polariton emission.

\subsection{Exciton-polariton lasing in nanocuboids.}


\begin{figure}[t!]
\centering
\center{\includegraphics[width=0.8\linewidth]{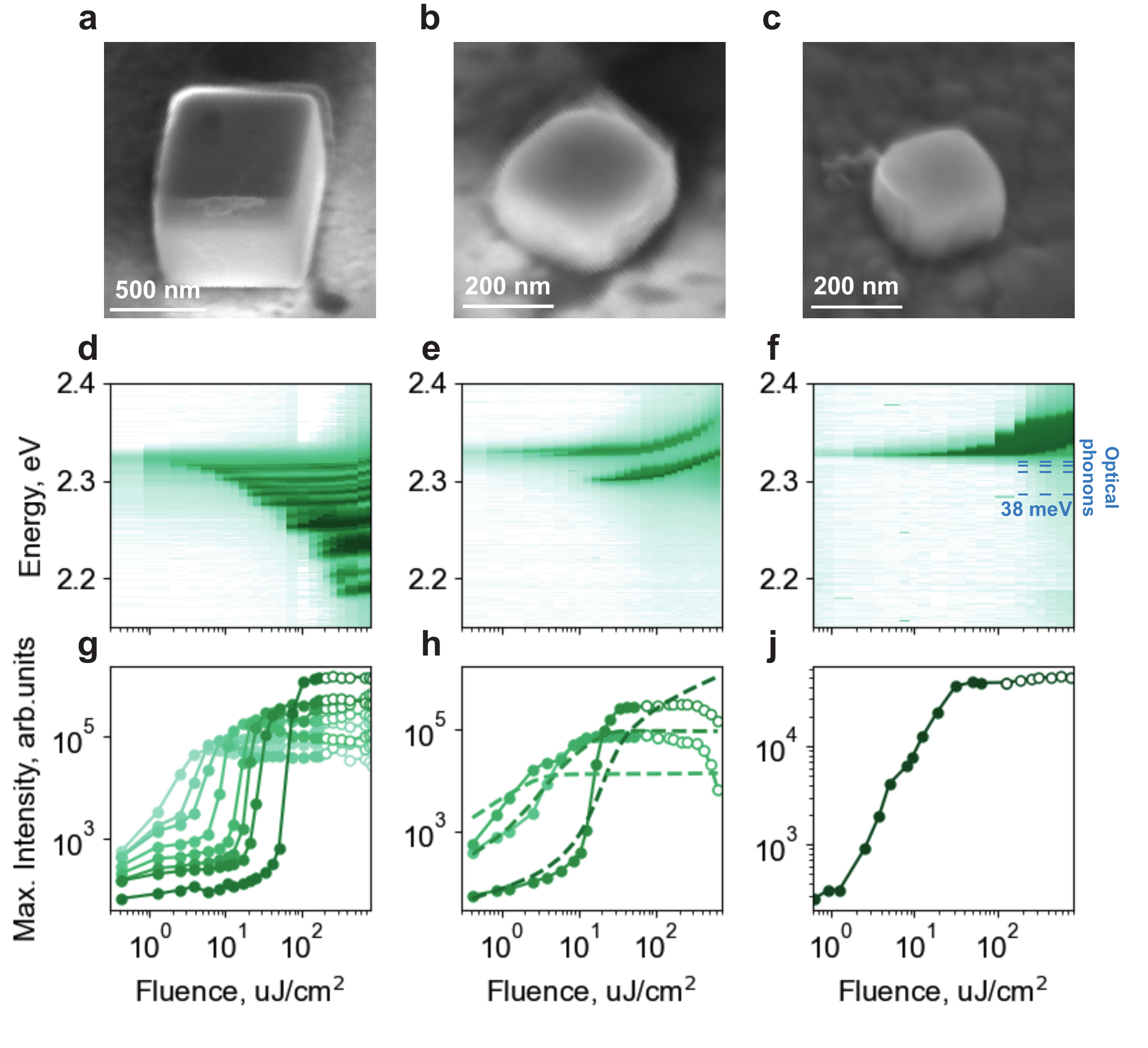}}
\caption{\textbf{Lasing emission measurements of the perovskite nanocuboids with different sizes.} (a-c) SEM pictures of studied perovskite nanocuboids with different geometry. The physical volume of the cuboids is 0.37, 0.02, and 0.007~$\mu m^3$ respectively. (d-f) Emission spectra as a function of the pump fluence obtained at 6~K for corresponding samples are shown below in (a-c). Intensive peaks in the spectra correspond to the lasing emission. Dashed blue lines in (f) correspond to the phonon energies shifted from the spectral center of the lasing mode.
(g-j) The intensity of the lasing peaks, shown in (d-f) as a function of the incident fluence. 
Dashed lines in (h) correspond to the results of the theoretical model of the polariton lasing and phonon relaxation in perovskite nanocuboid based on Eq.\ref{eq_model}.}
\label{fig3}
\end{figure}

The number of the polariton Mie-resonances in the perovskite nancuboids is determined by their sizes. Thus, we synthesize nanocuboids with different sizes during the study (See Methods and Fig. S7 in SI for the details), and choose the most relevant volumes (0.37 $\mu m^3$, 0.02 $\mu m^3$ and 0.007$\mu m^3$) which demonstrate multimode lasing, few-mode lasing and single-mode lasing regimes, respectively (See Fig. \ref{fig3}g-j).

With the increase in the pump fluence, the nanocuboid with the largest volume of 0.37 $\mu m^3$ (Fig. \ref{fig3}a) leads to multi-mode lasing accompanied by red broadening (Fig \ref{fig3}d), which corresponds to the similar polariton relaxation phenomenon, discussed above with 1D and 2D structures (Fig \ref{fig2}i). The large mode volume of the cuboid provides the numerous discrete Mie-states through which the stimulated polariton relaxation occurs. The mechanism for fluence-driven red broadening involves an increase in the number of phonons taking part in the polariton relaxation, resulting in the saturation of high-frequency modes and ascending of low-frequency ones. For 0.37 $\mu m^3$  cuboid, such a trend is vividly presented by the laser intensity dependency on fluence for every mode (Fig \ref{fig3}d).

However, the spectral dynamic drastically changes with a decrease in the cuboid size. We conduct the same measurements with 0.02 $\mu m^3$ cuboid, revealing lasing only on three modes according to experimental observations. The first mode appears upon excitation fluence around 1 $\mu$J/cm$^{2}$, pointing out the inversion-free regime. With an increase in the pump fluence, two additional channels of polariton relaxation towards the second and third Mie-modes appear at $\sim$~4~$\mu$J/cm$^{2}$ and $\sim$~17 $\mu$J/cm$^{2}$, respectively (Fig.~\ref{fig3}e). Further polariton thermalization is not observed, due to the large energy gap, exceeding phonon energies, or insufficient Q-factor of low-energy modes. Therefore, polaritons accumulate at three Mie-energy levels (E$_3$~> E$_2$~> E$_1$), resulting in polariton-polariton interaction through Coulomb potential and strong blueshift at the higher excitation fluence (F > 100~$\mu$J/cm$^{2}$)~\cite{masharin2022polaron}. This nonlinear regime is shown by empty circles in the mode intensity plot as a function of the pump fluence in Fig.~\ref{fig3}h. 

Notably, the pump fluence saturation for the higher energy mode (E$_2$) corresponds to the threshold fluence for the lower energy one (E$_1$). In accordance with the model based on coupled rate equations discussed above, the stimulated polariton relaxation provides the lasing at the E$_1$ level, when the nonlinear term of the relaxation to this level exceeds the rate of the annihilation. This in turn leads to the scenario when the particles at the E$_2$ level rather scatter to the lower state E$_1$ than radiatively annihilate at the E$_2$ level. Therefore, the lasing intensity of the second mode does not rise, when the stimulated relaxation channel for the first mode is opened. The hypothesis is supported by the results of calculations using the theoretical model described in the previous section, which are shown in Fig. \ref{fig3}h as dashed lines and well match with the experimental results. 


Encouraged by these findings we aim at the major challenge for the creation of the single-mode polariton laser having the smallest physical volume ever and operating upon an inversion-free regime. We study a cuboid with a physical volume around 0.007~$\mu m^3$ (Fig \ref{fig3}c) exhibiting completely suppressed polariton relaxation. It is established that laser generation occurs at an electric quadrupole Mie-mode at $\sim$~ 1 $\mu$J/cm$^{2}$ (Fig \ref{fig3}f,j). As expected, this cuboid demonstrates only the blueshift of laser emission because of the polariton accumulation at a single Mie-energy level.

\begin{figure}[t!]
\centering
\center{\includegraphics[width=0.9\linewidth]{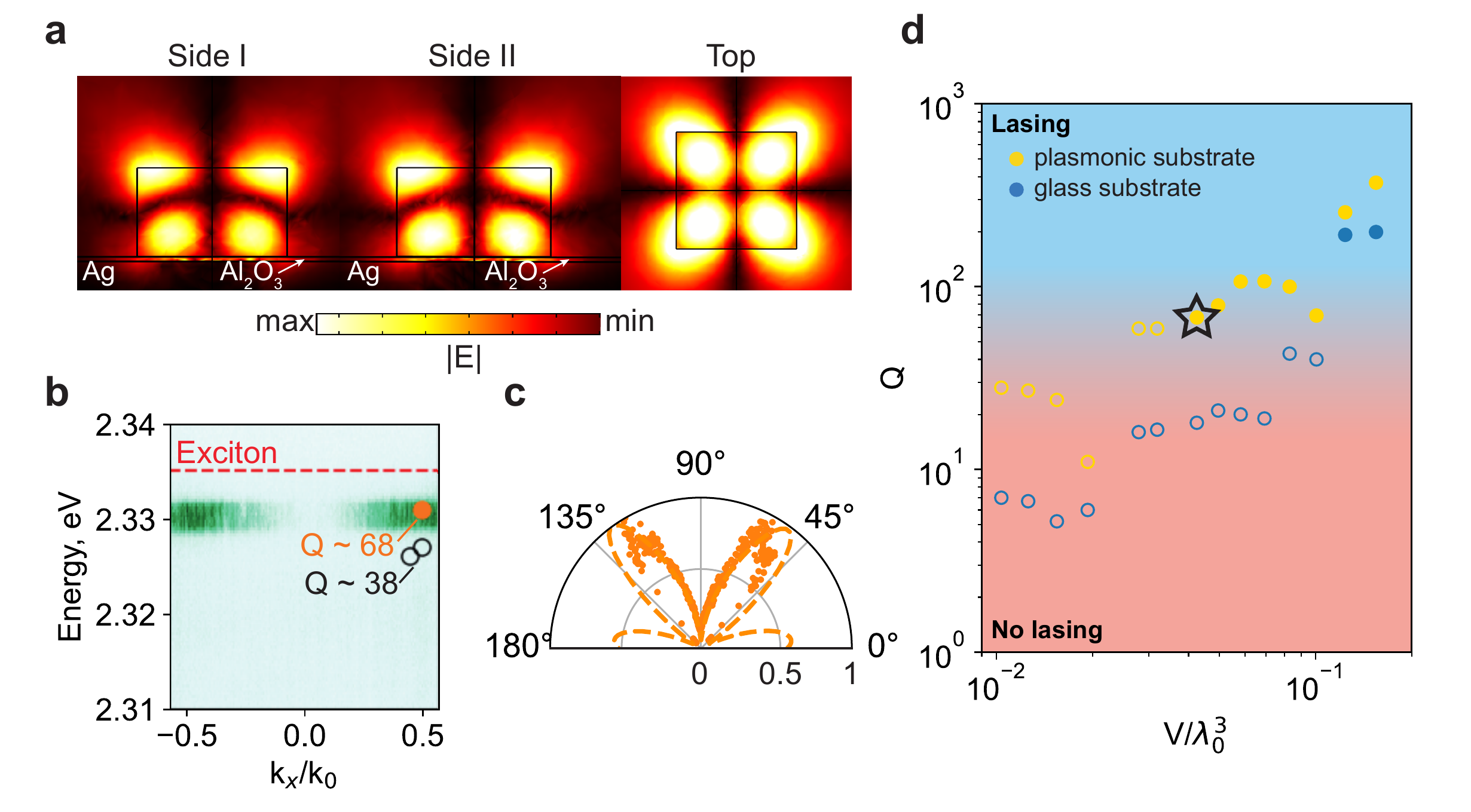}}
\caption{\textbf{Analysis of the Mie-mode observed in the smallest perovskite nanocuboid.} (a)-(c) Identification of the laser mode in cuboid with the smallest physical volume V = 0.007~$\mu m^3$ (0.04 $\lambda_{0}^3$, where $\lambda_0$~=~0.53~$\mu$m)  (SEM image in Fig.~\ref{fig3}c), providing a single mode lasing as shown in Fig.~\ref{fig3} f,j. (a) Electric field distribution of identified electric quadrupole (EQ) mode in three cross-sections. (b) Angle-resolved spectrum, obtained at fluence around 10 $\mu$J/cm$^{2}$. Presented laser line at 2.33 eV occurs at EQ resonance with Q~$\sim$~68 (shown by orange circle) with field localization shown in (a). Dashed red line corresponds to exciton energy level, while empty black circles point out to two more low-Q modes, presenting in this cuboid, but insufficient for lasing. (c) Angle-resolved spectrum in (b), plotted in polar coordinates (orange dots), and numerical directivity (dashed orange curve), obtained from far-field of EQ eigenmode (a). (d) The highest quality factor, Q, presenting in cuboid on metal-dielectric (yellow dots) and glass (blue dots) substrates versus normalized cuboid volume, $V/\lambda{_0}^3$. Here, the gradual transition defines the threshold level required for laser generation, obtained from semiempirical approach, whereas modes under threshold are noted by empty circles for clarity. The black star corresponds to the smallest laser achieved in this work, having a physical volume V = 0.007~$\mu m^3$   (0.047~$\lambda_{0}^3$, where $\lambda_0$~=~0.53~$\mu$m)  (SEM image in Fig.~\ref{fig3}c).}
\label{fig4}
\end{figure}


Dominating quadrupole nature of single lasing mode was confirmed by rigorous eigenmode analysis using Commercial Software COMSOL Multiphysics (See Methods and Fig. S10 in SI for details). Numerical analysis reveals that the smallest cuboid on a metal-dielectric substrate possesses a relatively high Q-factor for the resonance ($Q \sim 68$) hybrid Mie-mode at 2.33 eV, where the main contribution to mode field localization is governed by electric quadrupole with an azimuthal number m = 2 (further refereed as EQ) (Fig \ref{fig4}a).  We experimentally measure the angle-resolved emission of the smallest cuboid above the lasing threshold indicating that the radiation has a pronounced directivity within a solid angle defined by NA = 0.55 of a 50$\times$ objective lens, which at the normal direction the emission intensity tends to zero (Fig \ref{fig4}b). The obtained result is plotted in polar coordinates (orange dots in Fig \ref{fig4}c) and consistent with the directivity of EQ (orange dashed curves in Fig \ref{fig4}c) in Fig \ref{fig4}a. Therefore, both the spectral position and directivity of the calculated eigenmode agree well with the experimental observations, which points out the correct identification of the laser mode. A similar analysis is also performed for the middle-sized cuboid (Fig \ref{fig3}b), where mode with $Q_{1} \sim 59$ and $Q_{2} \sim 60$ correspond to non-degenerated (because of unequal sides) hybrid mode,  while the lower energy Mie-mode has $Q_{3} \sim 57.8$ with the major contribution from magnetic quadrupole with the azimuthal number m = 2 (Fig S3).

\begin{figure}[t!]
\centering
\center{\includegraphics[width=0.9\linewidth]{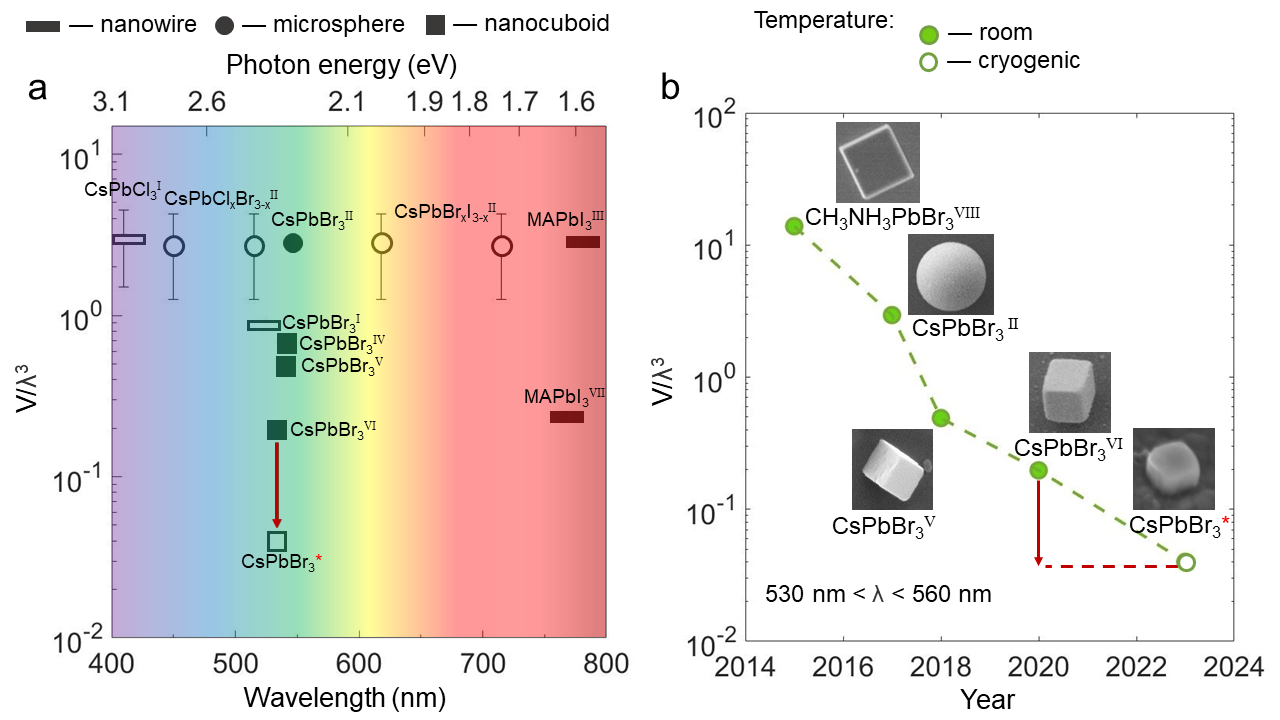}}
\caption{\textbf{Comparison of the smallest perovskite polariton nanolaser with previous reports.} (a) Reported perovskite laser designs (nanowire, microsphere and cuboids) having small normalized volumes, V/$\lambda^3$, with lasing wavelength over the whole VIS range at room (filled marker) and cryogenic (empty marker) temperatures. (b) Normalized volume versus year showing the status of green light emitting perovskite nanolasers miniaturization. Data were adapted from the following references:~\RomanNumeralCaps{1}~~\cite{jiang2018continuous};~\RomanNumeralCaps{2}~~\cite{tang2017single};~\RomanNumeralCaps{3}~~\cite{zhu2015lead};~\RomanNumeralCaps{4}~~\cite{cho2021submicrometer};~\RomanNumeralCaps{5}~~\cite{liu2018robust};~\RomanNumeralCaps{6}~~\cite{tiguntseva2020room};~\RomanNumeralCaps{7}~~\cite{yu2016organic};~\RomanNumeralCaps{8}~~\cite{liao2015perovskite}. The red arrow indicates progress achieved in this work.}
\label{fig5}
\end{figure}

Interestingly, pumping the nanocuboids with a physical volume smaller than 0.007~$\mu m^3$ (0.04 $\lambda_{0}^3$, where $\lambda_0$~=~0.53~$\mu$m) gives no lasing. To clarify this, the threshold radiative Q-factor required for lasing is identified to be around 52 on the basis of consistent data derived from experimentally observable laser modes and numerical modeling for corresponding eigenmodes. According to the defined threshold depicted in Figure~\ref{fig4}d as the gradual transition, cuboids on the plasmonic substrate (green circles) that have a value V/$\lambda_0^3$ less than 0.04 (signed with a star), apparently, are not capable of exhibiting lasing because of insufficient Q-factor. In turn, glass substrate does not afford laser generation in nanocuboids (blue circles) with V/$\lambda_0^3$ up to 0.12. Therefore, we promote the utilization of the metal-dielectric substrate providing the most compact laser semiconductor cavities with both efficient cooling and selective enhancement of certain Mie-modes (See Fig. S8 in SI).\cite{xifre2013mirror,sinev2016polarization} In the case of the studied EQ mode in the nanocuboid mentioned above, coupling to mirroring plasmonic substrate leads to an almost 7 times increase in Q-factor in comparison with the same cuboid on glass, where a substantial mode leakage takes place (Fig S10). Complementary experimental evidence of field localization enhancement provided by plasmonic substrate could be an observed acceleration of photoluminescence decay in the nanocuboid on the plasmonic substrate, as compared to that for similar cuboid on the glass substrate due to the Purcell effect (Fig S11).


To assess the contribution of our design to the concept of semiconductor laser medium miniaturization for the creation of building blocks in integrated photonic devices, we compare its physical and normalized volume as well as basic laser characteristics with those of other similar micro/nano lasers, operating in various spectral ranges (Table S1). As demonstrated in many compact perovskite designs, the color tunability by changing halogen at `X' site allows to easily obtain lasers in the whole visible spectral range with normalized volume, V/$\lambda^3$, close to 1 (Fig \ref{fig5}a). One of the most efficient strategies to push the compactness limit further is the integration of a perovskite cavity with a metal-dielectric substrate, which was demonstrated by \cite{yu2016organic,cho2021submicrometer}. Remarkable progress in laser miniaturization is obtained with transition from a nanowire towards a nanocuboid of the most structurally stable composition CsPbBr$_3$ (Fig \ref{fig5}a). Over the past decade, the size of “green” perovskite lasers has dropped down more than 2 orders of magnitude in terms of V/$\lambda^3$~ (Fig \ref{fig5}b). In this work, we integrate the smallest perovskite cavity (V/$\lambda^3~\sim$~0.04) with the metal-dielectric substrate yielding 5 times decrease in size as compared to the previous best result V/$\lambda^3~\sim$~0.2\cite{tiguntseva2020room}. Thus the produced single-mode Mie-polariton nanolaser sets a new record for miniature perovskite lasers (Fig \ref{fig5}a,b), providing them with the niche among the most compact semiconductor designs (Table S1).

\section*{Conclusion}

To conclude, we have proposed a strategy how to create ultra-small nanolasers and proved it experimentally. Namely, it is essential to create a cavity from the light-emitting material satisfying the following conditions: a high refractive index for modes localization in the nanocavity, a high binding energy and oscillator strength of the exciton to achieve polariton lasing, and a high quantum yield or luminescence to avoid overheating, as well as low phonon energies to prevent parasitic relaxation of exciton-polariton to lower states. We have supported this statement by rigorous experimental study of lead-bromide perovskite 0D cavity (nanocuboids) as well as for 1D and 2D cavities (nanowires and films). We have revealed that CsPbBr$_3$ perovskite is a promising material platform because of the high binding energy of exciton (35~meV), high refractive index near the exciton (>2.5), high quantum yield, and heavy ions resulting in low energies of optical phonons (<40~meV) being less than the intermodal spectral distance in a small Mie-resonant nanocavity. 

As a result of the optical design optimization, we have achieved exciton-polariton lasing in the nanocuboid laser based on the second order Mie resonance enhanced by a metallic substrate. The developed nanolaser possesses the smallest volume $(\lambda/2.7)^3$ among all perovskite lasers reported so far. Moreover, the nanolaser works at a very low threshold 1~$\mu$J/cm$^2$. Moreover, the developed method for colloidal synthesis of nanocuboid particles allows for the simple deposition of nanolasers, which makes it suitable for simple integration with various photonic circuits or conductive substrates, which is also promising for electrically driven nanolasers in the visible range.

\section*{Acknowledgements}
The work was supported by the Federal Program 'Priority 2030' and the Ministry of Science and Higher Education of the Russian Federation (Project 075-15-2021-589).
The authors are thankful to Nina Sheremet, Volodimir Sheremet, Hüseyin Bilge Yağcı, and Hammed Dehghanpour Baruj for assistance with the experiments at Bilkent University UNAM, and Mikhail Baranov for assistance with the experiments at ITMO University. The authors thank Mr. Ivan Pustovit for assistance in graphic design.






\section*{Methods}

\subsection*{Film synthesis}

MAPbBr$_3$ thin film was synthesized by the spin-coating method \cite{jeon2014solvent}. Perovskite solution was prepared in the nitrogen dry box by mixing 56.0~mg of methylammonium bromide (MABr) from GreatCell Solar and 183.5~mg of Lead(II) bromide (PbBr$_2$) from TCI in the glass vial. Salts were dissolved in 1~mL of DMF:DMSO solvent mixture in the ratio 3:1. Obtained  MAPbBr$_3$ solution with 0.5M molarity was stirred for 1 day at 27 C$^{\circ}$. 

We used SiO$_2$ substrates (12.5$\times$12.5 mm), which were washed with sonication in the deionized water, acetone, and 2-propanol for 10~min consecutively, and afterward cleaned in an oxygen plasma cleaner for 10~min. Prepared substrate transfer to the dry nitrogen glovebox for further spin-coating. 

First, the substrate was located on the spin-coater and fixed by a vacuum pump, then 30~$\mu$L: of the MAPbBr$_3$ solution was deposited on the top of the substrate, and after the substrate was spun at 3,000~rpm for 40~s. At the 25 s 300~$\mu$L of toluene was dripped on the top of the rotating substrate to rapidly wash out the solvent and trigger the crystallization. When rotation stoped substrates with perovskite film were annealed at 90 $^{\circ}$C for 10 min. 

\subsection*{Perovskite nanowire synthesis}

Perovskite nanowire synthesis was performed by the developed solution-engineering method.\cite{pushkarev2018few} First, CsPbBr$_3$ solution was prepared: 110 mg of PbBr$_2$ and 63 mg of CsBr from TCI were dissolved in 3 mL of DMSO by shaking for 10 min to afford a clear solution in a dry nitrogen glovebox atmosphere. Glass substrates were cleaned at ambient conditions with chromium oxide paste and rinsed with distilled water to ensure the hydrophobic surface. The prepared perovskite solution was taken out from the glovebox, exposed to the 30\% humid air, and kept in the closed vial for 15 min. Then 20 $\mu$L of the solution was deposited on the rotating substrate at 2,000 rpm for 30 s. After the sample was put in a small plastic Petri dish bottom (35 x 9 mm$^2$) placed in a bigger glass one (80 x 15 mm$^2$). The whole system was heated up to 50 $^{\circ}$C and 200 $\mu$L of IPA$\cdot$H$_2$O azeotrope was poured into the glass Petri and all the system was capped with a glass Petri top. At the final stage, the sample was annealed in the system at 50 $^{\circ}$C for 6 min in the presence of 2-propanol-water azeotropic vapor.

\subsection*{Perovskite nanocube synthesis}

The following protocol reproducing a hot-injection method~\cite{protesescu2015nanocrystals} adapted to synthesis of perovskite Mie-resonance nanocuboids was employed. First, cesium oleate (CsOA) in 1-octadecene (ODE) 0.125 M solution was prepared. For this, 2.5 mmol of Cs$_2$CO$_3$ was loaded in a flask containing 40 mL of ODE and stirred (1,000 rpm) at 140 $^{\circ}$C in a N$_2$-filled glovebox for 1 h followed by its reaction with 5 mmol of oleic acid (OA) giving a clear solution that turns to be milky at room temperature. The solution was stored in the glovebox and preheated before the hot-injection procedure. Second, 0.2 mmol of PbBr$_2$ was stirred (1,000 rpm) in 5 mL of ODE in a vacuumized two-neck round-bottom flask (50 mL) at 120 $^{\circ}$C for 1 h. Thereafter, the flask was filled in with N$_2$ gas, heated up to 140 $^{\circ}$C, and 0.4 mmol of oleylamine (OLAm) and 0.4 mmol of OA were consequently added dropwise to completely dissolve lead bromide powder. Then, a clear solution was heated up to 180 $^{\circ}$C followed by rapid injection of 0.05 mmol of CsOA in ODE solution. The reaction mixture was stirred at 180 $^{\circ}$C for 10 min and was quenched in an ice bath afterwards. The obtained suspension was centrifuged at 500 rpm for 5 min. A supernatant containing small perovskite nanocrystals was discarded. A precipitate was redispersed in 10 mL of toluene. The suspension was left for 20 min to let the Mie-resonance nanocuboids settle out. Finally, the supernatant was pipetted out and the sediment was redispersed in 10 mL of n-hexane for further utilization.

Metal-dielectric Al$_2$O$_3$/Ag/Si substrates were fabricated by using the following approach. First, Ag film with a thickness of 50 nm was deposited on Si substrates at 0.2 A/s rate by thermo-resistive evaporation in a vacuum chamber (Nanovak) at 2x10$^-6$ torr pressure. Thereafter, the resultant Ag/Si samples were coated with 6 nm layer of amorphous aluminum oxide produced by atomic layer deposition (ALD) using Al$_2$(CH$_3$)$_6$ (trimethylaluminium) and H$_2$O as precursors. 60-cycle deposition was conducted at 300 $^{\circ}$C and coating rate 1.05 A/cycle on Cambridge NanoTech Savannah Atomic Layer Deposition System. The AFM image of the substrate surface was obtained on a AIST SmartSPM 1000 atomic force microscopy (See Fig. S6 in SI for details).

Suspension of CsPbBr$_3$ nanocuboids in n-hexane was dropcasted on metal-dielectric and glass substrates for carrying out optical experiments.The morphology and size of the nanocuboids were studied using Zeiss Neon 40 and Zeiss Merlin scanning electron microscopes. 

\subsection*{Optical measurements}
Pump-dependent emission spectra with angular distribution was obtained by a back-focal-plane setup with a slit spectrometer coupled to a liquid-nitrogen-cooled imaging CCD camera (Princeton Instruments SP2500+PyLoN). Femtosecond (fs) laser (Pharos, Light Conversion) coupled with a broad-bandwidth optical parametric amplifier (Orpheus-F, Light Conversion) was used as a pump for the PL excitation at the wavelength of 490 nm with a repetition rate of 100 kHz. Both for the focusing of the laser beam and for collecting emission signal infinity-corrected objective (Mitutoyo 50x NIR HR with NA = 0.65) was used.  Samples were placed in a closed-cycle helium cryostat (Advanced Research Systems) and maintained at a controllable temperature in the range of 7-300~K. The spatial filter was used to get rid of parasitic signals from the neighborhood nanocrystals and reflections from the optical elements in the setup.

Time-resolved measurements at room and cryogenic temperatures were carried out with using a PicoHarp 300 TCSPC module (PicoQuant). A femtosecond laser (Pharos, Light Conversion) coupled with a broad-bandwidth optical parametric amplifier (Orpheus-F, Light Conversion) was used as a pump source ($\lambda_{ex}$~= 490 nm, RR = 100 kHz, $\tau$~$\approx$~220 fs). The laser beam was focused on the sample surface, placed in a closed-cycle helium cryostat, at normal incidence by an infinity-corrected objective (Mitutoyo 50x NIR HR NA = 0.65) providing the sample with uniform irradiation (Gaussian distribution with FWHM of $\approx$~10~$\mu$m). Spontaneous PL emission was collected by the same objective and sent towards a photon counting detector modules (photon timing resolution of 50 ps (FWHM), PDM Series, Micro Photon Devices). To get rid of residual excitation light the longpass filter with cut-on wavelength 500 nm (FELH0500, Thorlabs) was used. The spatial filter was used to get rid of parasitic signals from the neighborhood nanocrystals and reflections from the optical elements in the setup. 

For the angle-resolved measurements of the perovskite film guided mode below the light line (k$_x$/k$_0$ > 1) ZnSe solid immersion lens (SIL) was used.\cite{permyakov2021probing} The flat side of the SIL was placed close to the thin film to the distances in the order of 100 nm by hand made setup with piezo positioner. Infinity-corrected objective (Mitutoyo 100x HR VIS 0.9 NA) was focused through the SIL. The resulting NA of the system becomes equal to NA = $0.9 \cdot n_{ZnSe} \approx 2.4$ in the studied spectral range. Collected signal from the SIL setup was transmitted to the mentioned back-focal-plane setup to get angle-resolved spectra. For the reflectance spectra halogen lamp was used, and for the pump-dependent PL spectra fs laser (Pharos, Light Conversion) at 515 nm with a 100 kHz repetition rate was used.

\subsection*{Fitting of guided polariton dispersion in thin film}

To fit polariton branch we take particular spectrum for each measured available k$_x$/k$_0$ in the range from 1.65 to 1.9, shown in Fig \ref{fig2}c and fit experimental data by the Lorentz peak function. Taking center of the Lorentz peak function as a function of k$_x$/k$_0$ we obtain the experimental lower polariton dispersion. 

In the current design there is no upper polariton branch because of huge absorption above exciton resonance, therefore we can estimate Rabi splitting and light-matter coupling coefficient by two coupled oscillator model, using only lower polariton branch \cite{hopfield1958theory}. According to the model lower polariton can be described as: 

\begin{equation}
    E_{LP}(k) = \frac{\widetilde{E}_{X} + \widetilde{E}_{C}(k)}{2} - \frac{1}{2}\sqrt{(\widetilde{E}_{C}(k) - \widetilde{E}_{X})^2 + 4 g_0^2}, \label{CoupledOscillator}
\end{equation}
where $E_{LP}(k)$ is lower polariton dispersion, $\widetilde{E}_{X}$ and $\widetilde{E}_{C}(k)$ is complex energy of exciton and uncoupled cavity photon respectively,  $g_0$ is the light-matter coupling strength. Exciton level was estimated from PL and reflection spectra, uncoupled cavity photon dispersion is linearly approximated in the spectral range far from exciton. By optimizing the parameter $\Omega_0$ we fit experimental dispersion by the current model. Resulting real part of $E_{LP}$ is shown in Fig.~\ref{fig2}b,c 

\subsection*{Mode analysis}
Mode analysis of lasing in CsPbBr$_3$ cuboids at cryogenic temperature was performed in the frequency domain module of COMSOL Multiphysics software package. Eigenfrequency study was used to carry out the eigenmode calculation for cavity with real geometrical parameters estimated from SEM image, radiation patterns of eigenmodes were calculated with using far-field domains. The surrounding area of the perovskite nanocuboid was simulated as a hemisphere of air (n = 1) with a radius of 1 $\mu$m with the outer perfectly matched layer. Following material parameters were employed: complex refractive index of CsPbBr$_3$ at $\sim$~6~K, calculated by analytical approach (See Section S7 of SI for more details), whereas optical constants for metal-dielectric substrate Ag, Al$_2$O$_3$ were taken from~ \cite{mcpeak2015plasmonic, boidin2016pulsed}, respectively. A constant refractive index of CsPbBr$_3$, n~=~2.3, was used for carrying out the comparison with the case when the refractive index is constant for all light wavelengths.  

\bibliography{main}

\section*{Author contributions statement}
M.A.M. provided the thin film fabrication, conducted the optical experiments, developed the concept, and analysed experimental data. 
D.K. synthesized perovskite nanocuboids and nanoplates, provided experiments with nanocuboids, characterized the plasmonic substrate, and analysed the experimental and theoretic results. 
V.I.K. provided the experimental setup for SIL angle-resolved spectroscopy measurement and helped with experiments with it.
D.I.M. provided the synthesis of the perovskite nanowhiskers and participated in the experiments with them.
A.D.U. provided the theoretical model for multimode analysis of perovskite nanocuboids and simulate the scattering radiation pattern. 
A.D.D. provided the theoretical estimation of the CsPbBr$_3$ refractive index at 6~K.
V.A.S. provided the theoretical model of stimulated polariton relaxation.
A.P.P. supervised the fabrication process.
F.I. participated in the SEM measurements and plasmonic substrates fabrication.
I.V.I. and I.A.S. supervised the theoretical part of the work.
H.V.D. supervised the experimental part of the work. 
A.K.M. suggest the main concept of the polariton-assisted stimulated emission and supervised the experimental work. 
S.V.M. managed the whole project. 
All authors extensively discussed the results and participated in editing the manuscript.

\begin{figure}[t!]
\centering
\center{\includegraphics[width=0.6\linewidth]{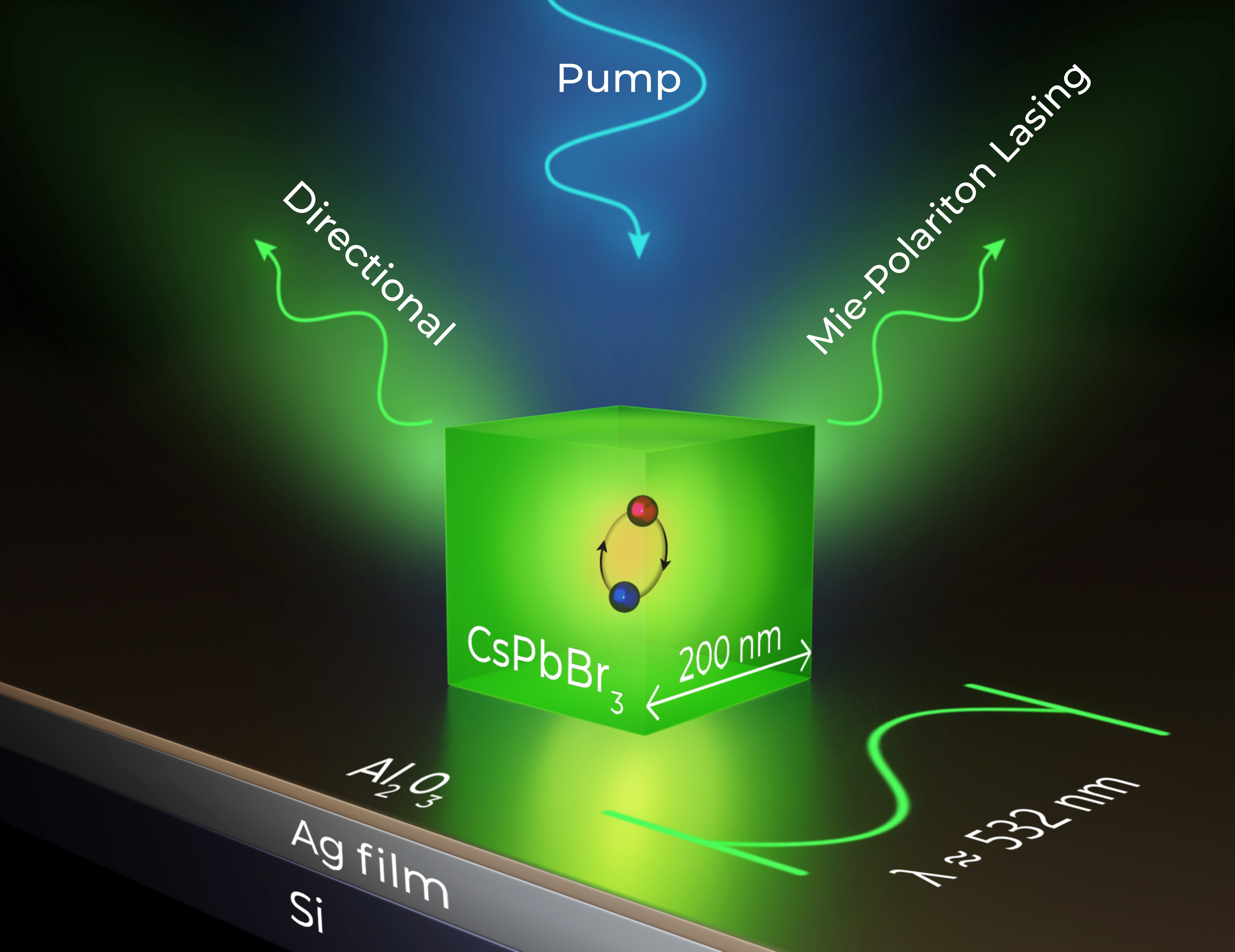}}
\caption{ToC figure}
\label{figTOC}
\end{figure}

\newpage

\renewcommand{\thefigure}{S\arabic{figure}}
\renewcommand{\theequation}{S\arabic{equation}}

\section*{
Supplementary materials: Polariton lasing in Mie-resonant perovskite nanocavity
}

\section*{Section S1. Experimental measurement of ASE from perovskite thin film under the light cone}

\begin{figure}
\centering
\renewcommand\thefigure{S1}
\center{\includegraphics[width=1\linewidth]{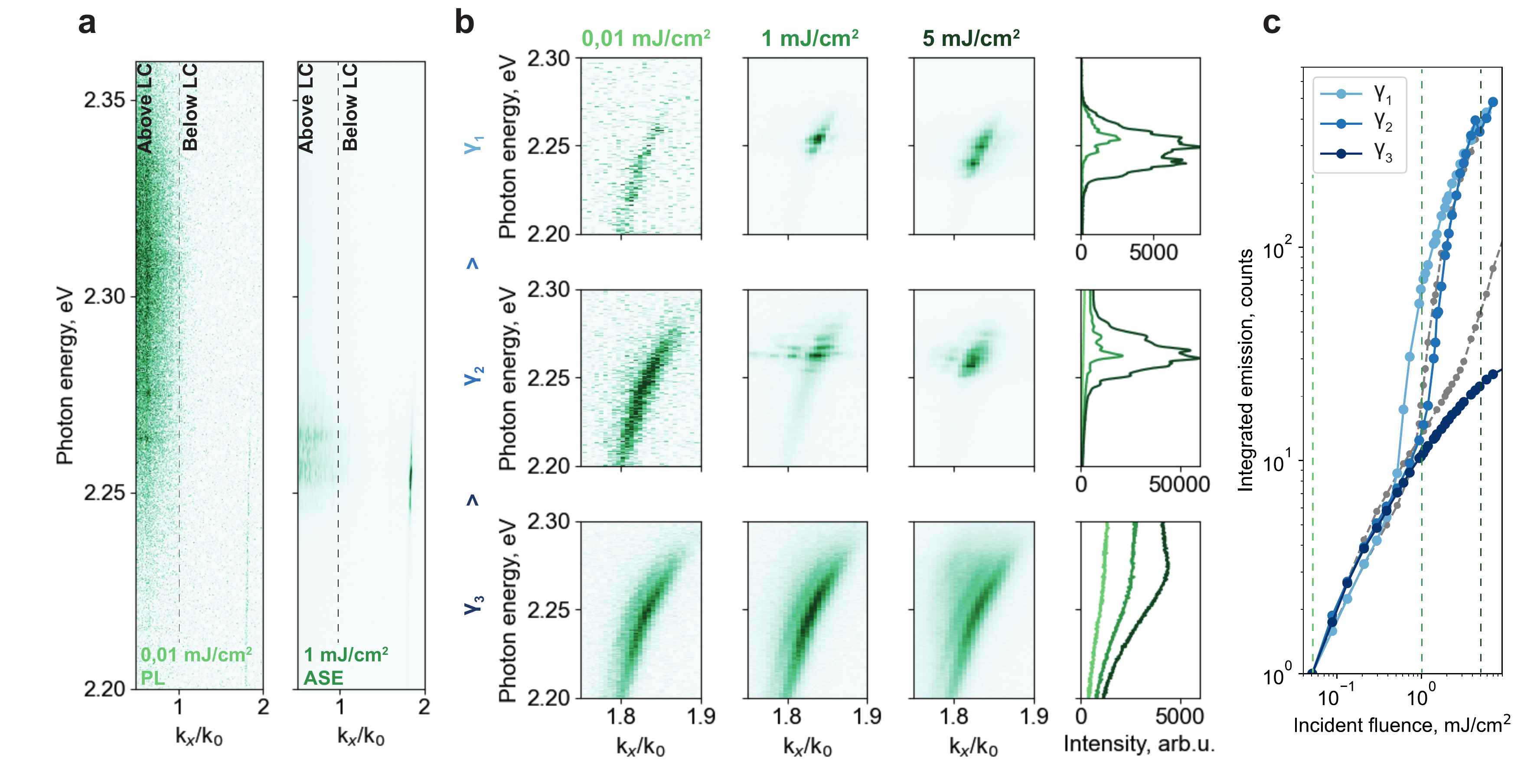}}
\caption{(a) Angle-resolved emission spectra of guided polariton, obtained at room temperature under femtosecond pump with fluence around 0.01 mJ/cm$^2$ and 1 mJ/cm$^2$ in the regime of low radiative losses  (b) Obtained spectra with different pump fluence and SIL distance to the sample. Last dictates mode losess $\gamma$. On the figure the value of the ASE red brodaning for high pump fluences decrease with increasing  of mode losses. For the highest mode losses ASE was not observed. (c) Integrated emission signal as a function of incident fluence for different waveguide mode losses. Colored dots correspond to the optical losses, shown in b, grey dots correspond to the measurements, which are not presented in the text.}
\label{fig_ASE_vs_gamma}
\end{figure}

As we discuss in the main text, perovskite polycrystalline thin film is able to support waveguide modes, which can be directly measured by angle-resolved spectroscopy with SIL. Angle-resolved emission spectra, measured under femtosecond laser pump at 515~nm at pump fluences around 0.01 mJ/cm$^2$ and 1 mJ/cm$^2$ in the regime of low coupling of SIL (when the waveguide is located far from the SIL). The waveguide mode is shown in Fig.~\ref{fig_ASE_vs_gamma}a with k$_x$/k$_0 \approx$ 1.8. The low coupling of the SIL provides low radiative optical losses of the waveguide mode. In the case of low pump fluence, we observe a broad intensive PL signal above the light cone and a low-intensive waveguide mode below the light cone. According to our assumptions, in this regime, we measure exciton PL with thermalized polaritons over the branch. At the pump fluences around 1 mJ/cm$^2$ above the light cone, ASE emission appears in the spectral region around 2.26 eV. At the same time below the light cone, we also observe enhancing of the intensity in the same spectral region, corresponding to the ASE. Moreover, the intensity, measured below the light cone strongly exceeds the intensity above. It can be explained as the origin of ASE came from the guided polariton and then it is scattered via defects and film grains, outcoupling to the free space. 

In the experiment we also change the distance between SIL and perovskite thin film, changing the SIL coupling efficiency and hence radiative losses of waveguide mode (Fig.~\ref{fig_ASE_vs_gamma}b). In the case of the lowest radiative losses $\gamma_1$ we observe pronounced red broadening of ASE with increasing pump fluence. When we decrease the distance between SIL and film with increased radiative losses $\gamma_2$ the red broadening is suppressed. In the regime of the SIL overcoupling with highest optical losses $\gamma_3$ we do not observe any ASE. The phenomenon can be explained in the terms of the polariton lifetimes. When we increase radiative losses of guided mode, polaritons recombine either scatter to the lower energies. If losses are too high, polariton concentration cannot achive nonlinear regime because of fast recombination, which we observe in the case of $\gamma_3$.

We also study light-light curves for different guided polariton radiative losses shown in Fig~\ref{fig_ASE_vs_gamma}c. The ASE provides pronounced threshold behavior in the case of low optical losses. Moreover, the ASE threshold is increasing with increasing waveguide optical losses. As there exists threshold concentration of polaritons for stimulation, increasing losses leads to the requirement of the higher pump fluence to achieve nonlinear regime. Eventually, the ASE threshold can be larger than the degradation threshold and we will not able to see any ASE, as we demonstrate in the case of $\gamma_3$ in Fig~\ref{fig_ASE_vs_gamma}c.   

\section*{Section S2. Interference pattern on angle-resolved emission spectra from perovskite nanowires}
\begin{figure}
\centering
\renewcommand\thefigure{S2}
\center{\includegraphics[width=0.8\linewidth]{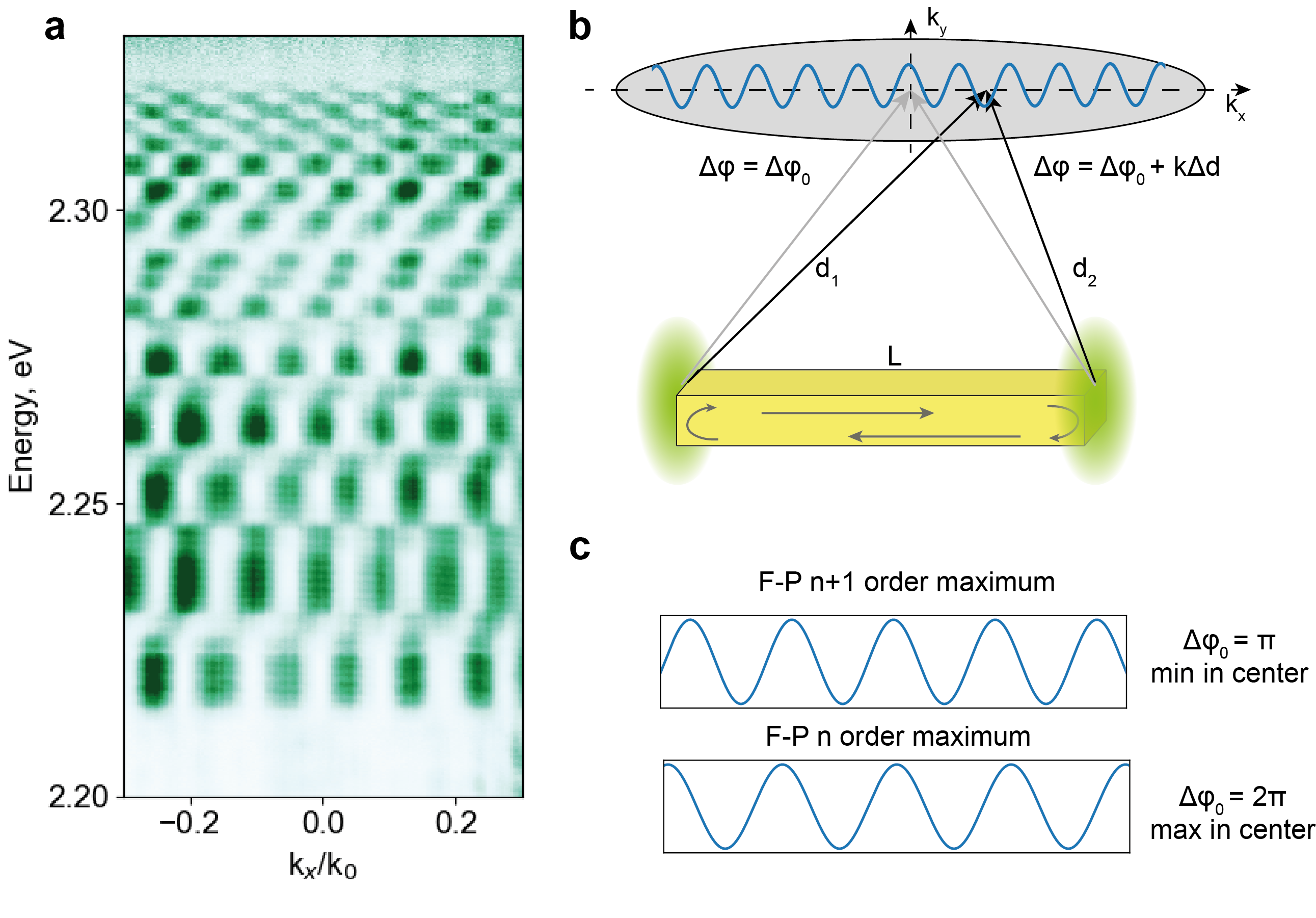}}
\caption{(a) Angle-resolved emission spectrum, measured at 850 $\mu$J/cm$^2$. Data is normalized to the integrated spectrum at each k$_x$/k$_0$ to better visualization of the pattern. (b) Sketch, visualizing the nature of interference pattern, obtained at panel a; (c) Vizualization of the field distribution in the nanowire of two neighboring F-P resonance. Showed phase difference $\Delta \phi_0$ is between two different edges of nanowire for both resonances}
\label{figS_nanowire}
\end{figure}

Angle-resolved pump-dependent emission measurements from perovskite nanowires are performed in the case of the nanowire axis co-directed with an angle-resolved $x$-axis. To study the interference pattern we normalized angle-resolved spectra from Figs.~2g-i in the main text to the integrated over the k$_x$/k$_0$ spectra to make it more pronounced (Fig.~\ref{figS_nanowire}a). At the high pump fluence in the multilasing regime, the emission becomes coherent in the nanowire and outcouples through the edges. The pattern appears due to the interference of the coherent emission from the opposite edges of the nanowire on the BFP (Fig.~\ref{figS_nanowire}b). 

The number of the interfered maxima over k$_x$/k$_0$ depends on the distance between coherent light sources, or in other words, the physical nanowire length. It should be noticed, that pattern changes the phase over the $\pi$ with each next order. It can be explained either an even or an odd number of half-length is suited to the nanowire optical length. The parity changes with every next order and therefore the initial phase difference $\Delta \phi_0$ also drop from $\pi + 2\pi n$ to $2\pi n$ or vice versa (Fig.~\ref{figS_nanowire}c). The observed effect helps to distinguish several F-P resonances from each other near the exciton, where it is located very close to each other and when the intensity slightly changes from one resonance to another. By extracting the F-P resonances close to the exciton in the following way, we calculated the free spectral range of the F-P resonances and estimated the group refractive index, shown in Fig.~2j in the main text.

\section*{Section S3. Radiation pattern of Mie-polariton lasing emission from the nanocubes}

\begin{figure}
\centering
\renewcommand\thefigure{S3}
\center{\includegraphics[width=0.8\linewidth]{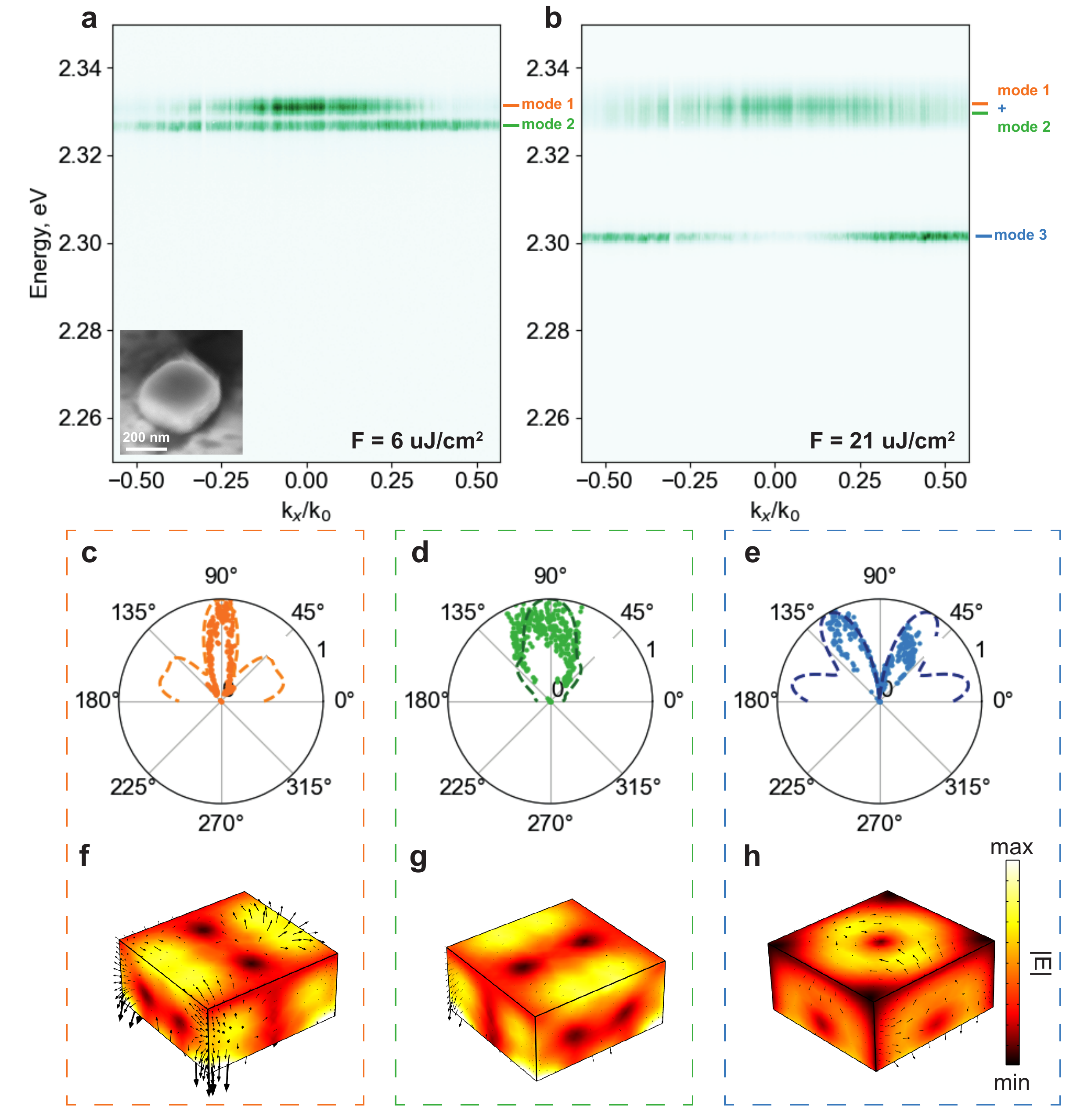}}
\caption{(a,b) Angle-resolved emission spectrum of  0.02 $\mu m^3$ cuboid upon 6 and 21 $\mu$J/cm$^2$ pump fluence. (c-e) Radiation distribution of the lasing modes measured experimentally, shown as dots, and simulated far-field pattern, shown as dashed lines. (f-g) Corresponding electrical field distribution of the calculated eigenmodes.}
\label{figS_cube_rad}
\end{figure}

Pump-dependent emission measurements shown in the main text were done also with angle resolution. Angle-resolved emission spectrum measured from 0.02 $\mu m^3$ cuboid upon 6 $\mu$J/cm$^2$ is shown in Fig.~\ref{figS_cube_rad}a. It is shown two lasing peaks, noted as "Mode 1" and "Mode 2", which are separated on the wavelength. Third mode (“Mode 3”) appears at around 2.3 eV at higher fluence as illustrated in Fig.~\ref{figS_cube_rad}b, whereas the first two modes become closer to each other and cannot be clearly distinguished.  To make it more clear we plotted the section over the wavelength peak centers in the radiation plot demonstrated in Fig.~\ref{figS_nanowire}b.  Polar plots for modes 1 (orange dots), 2 (green dots), and 3 (blue dots) have different radiation patterns (Figs.~\ref{figS_cube_rad}c-e). According to numerical eigenmode analysis “Mode 1” (with $Q_{1}\sim 59$ ) and “Mode2” ($Q_{2} \sim 60$) correspond to non-degenerated (because of unequal sides) hybrid mode consisting of mixture of spherical harmonics\cite{gladyshev2020symmetry} involving magnetic octupole with azimuthal number m = ~$\pm$~1, while the lower energy Mie-mode has $Q_{3} \sim 57.8$ with the major contribution from magnetic quadrupole with azimuthal number m = 2 (Fig.~ \ref{figS_cube_rad}). Spatial electrical field distribution of each eigenmode is presented in Figs.~\ref{figS_cube_rad}f-h, while calculated far-field directivity is shown in Figs.~\ref{figS_cube_rad}c-e by dashed curves matching well the experimental results. 

\section*{Section S4. Experimental observation of the Mott transition in the perovskite nanocubes}

In the previous works, it was suggested, that lasing in the CsPbBr$_3$ nanowires appears above the Mott transition and originated from the electron-hole plasma and plasmonic resonances. \cite{schlaus2019lasing} Authors sustained their claim by the ultrafast emission spectroscopy, where lasing peaks in the spectra shifts to the red region with time, which points to the dependence of the peak position with carrier concentration. However, the same result can be explained by the polariton nature, mentioned in the main text. 

\begin{figure}
\centering
\renewcommand\thefigure{S4}
\center{\includegraphics[width=0.8\linewidth]{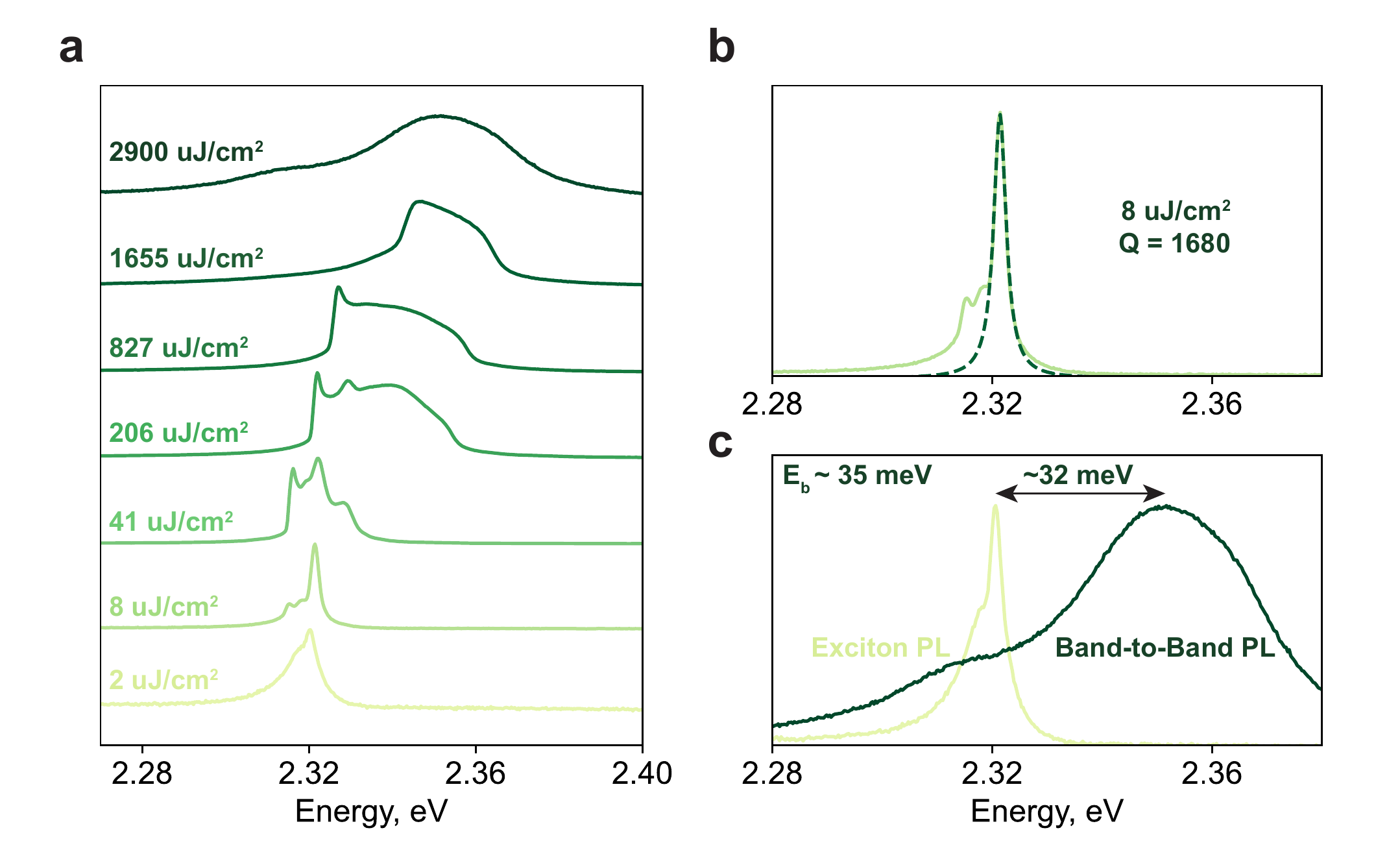}}
\caption{(a) Integrated emission, obtained under femtosecond pump at 6 K with increasing of incident fluence. (b) Fit of the lasing peak at 8 $\mu$J/cm$^2$ with calculated quality factor (c) Normalized integrated spectra, obtained at 2 and 2900 $\mu$J/cm$^2$. First corresponds to exciton PL and the second to band-to-band PL. The difference between these two spectra are close to the exciton binding energy in this material. }
\label{figS_Mott}
\end{figure}

To check the lasing origin in our experiments we pump our nanocubes at the highest pump fluences at 6 K to experimentally achieve the Mott transition (Fig.~\ref{figS_Mott}a). As was already discussed in the main text, we observe a few-lasing peak regime from the nanocube at the low pump. The quality factor of the lasing mode achieves 1680 (Fig.~\ref{figS_Mott}b) at 8 $\mu$J/cm$^2$. With increasing the pump fluence lasing peaks shift and broad to the blue region. At the fluence around 1000 $\mu$J/cm$^2$, broad background PL appears, where stimulated emission still exists. Finally, at fluences above 2500 $\mu$J/cm$^2$ lasing stimulated emission disappears and we observe a broad PL spectrum. The spectral center of the broad PL spectrum is shifted from the initial PL center by the value around 32 meV. As the exciton binding energy is estimated to be around 35 meV \cite{khmelevskaia2021excitonic} we attribute the initial PL to the exciton recombination and PL obtained at the high pump fluence to the band-to-band recombination. In other words, we achieve the Mott transition at the pump fluences around 1000 $\mu$J/cm$^2$, which is described in a broad PL background, but the Coulomb correlations between electrons and holes still exist and the polariton lasing can still be maintained, but with lower intensity. At the higher fluences, even the correlation dissapears with the lasing emission and we obtain clear band-to-band PL.  It also should be noted, that according to the experiment the process is reversible, which means, that we are below the degradation threshold for the particular perovskite nanocube. According to the estimation,  Mott transition fluences are close to the estimated critical Mott concentration, which fits the known experimental data. For more details see Section \red{S5}. This data proves, that the lasing emission in our case originated from the exciton-polariton nature of either electron-hole plasma.

\section*{Section S5 Estimation of the excited exciton concentration in the nanocubes}
In order to compare experimentally achieved Mott transition with literature data we estimate excited carrier concentration in our experiments.
In the simplest estimations absorbed optical power intensity can be estimated as

\begin{equation}
    I_{abs} = I_{inc} T \left(1 - e^{-\kappa k_0 d}\right)
\end{equation}
where $I_{inc}$ is incident power, T is transmittance coefficient, $\kappa$ is the imaginary part of the refractive index at the pump wavelength, $k_0$ is incident pump k-vector and $d$ is the height of the nanocube. In the pulsed pump regime it can be rewritten in the carrier number notations: 

\begin{equation}
    N_{abs} = N_{inc} T \left(1 - e^{-\kappa k_0 d}\right)
\end{equation}

The absorbed carrier concentration $n_{abs} = N_{abs} / S d$, where S is the scale area of the incident pulse. And $N_{inc} =  F_{inc}S / E_{ph}$, where $F_{inc}$ is incident fluence and $E_{ph} = h c/ \lambda$ is the energy of the single photon at the pump wavelength. Resulting carrier concentration can be estimated as:

\begin{equation}
    n_{abs} = T \frac{F_{inc}\left( 1 - e^{-\kappa k_0 d} \right)}{E_{ph} d}
\end{equation}

According to the previous studies, critical exciton Mott transition density is estimated to be equal $n_M = 1.0 \cdot 10^{19}$.\cite{su2021perovskite} Based on the current approximations in our experiment with a 490 nm pulsed pump, such concentrations are reachable at the fluences around 300 $\mu$J/cm$^2$. 
Deviations from the experimental observation can be explained by overestimation of the excited exciton concentration, which appeared from overestimated absorption coefficient at 6 K, Gaussian distribution of the pump spot profile, and optical effects due to the substrate, and nanocube faces. 
Also, it should be noted, that estimated Mott concentration does not take into account polaron effects in the material \cite{evans2018competition}, which increases critical Mott concentration.\cite{masharin2022polaron} 
Nevertheless, the order of the estimated excited exciton concentration and Mott concentration is well agreed with experimental observations and prove, that observed lasing and ASE regimes originated from exciton-polariton nature.

\section*{Section S6 Optical properties of CsPbBr$_3$ cuboids at room temperature}

\begin{figure}[h!]
\centering
\renewcommand\thefigure{S5}
\center{\includegraphics[width=0.8\linewidth]{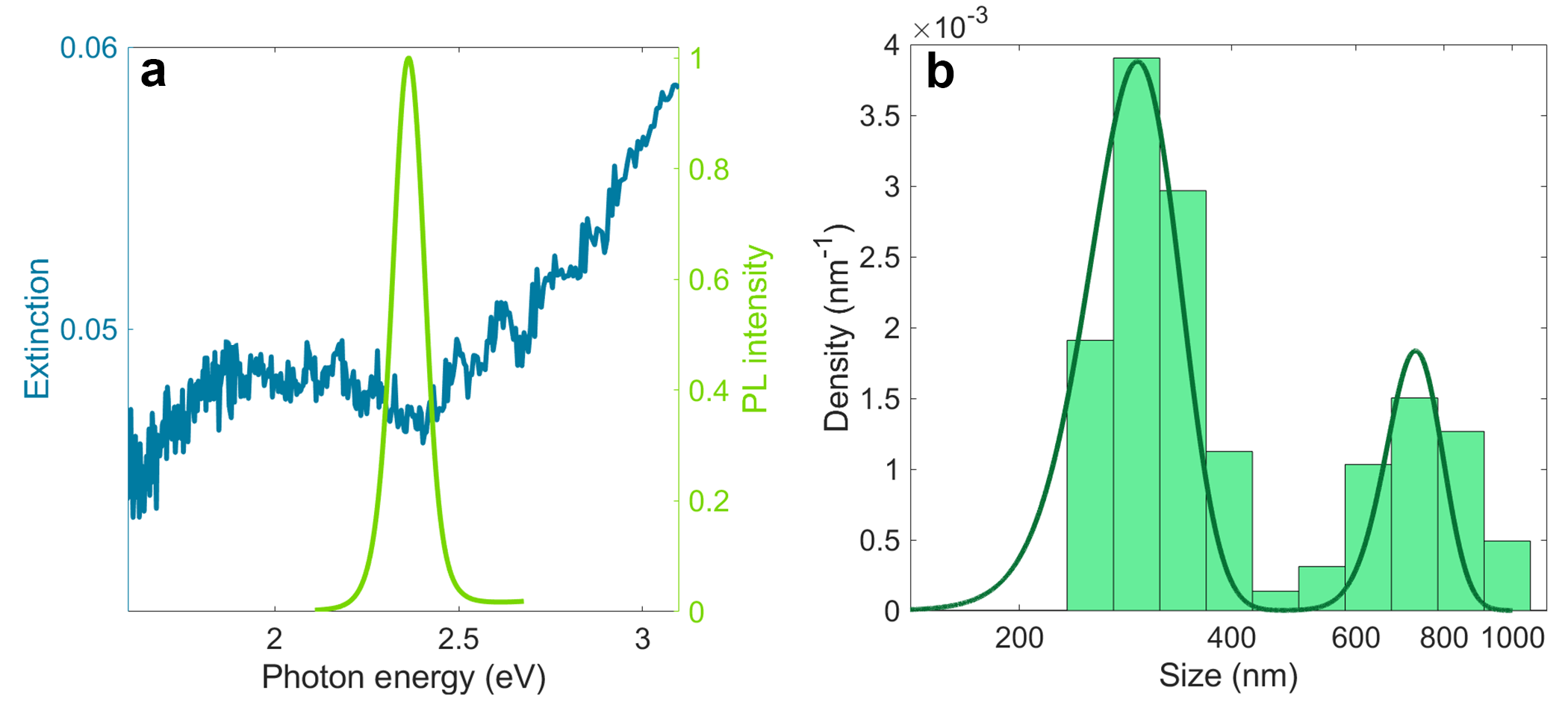}}
\caption{(a) Photoluminescense and extinction spectra of nanocuboids suspension in n-hexane. (b) Size  distribution  histogram  of nanocuboids suspension in n-hexane.}
\label{Solut}
\end{figure}

\begin{figure}
\centering
\renewcommand\thefigure{S6}
\center{\includegraphics[width=0.8\linewidth]{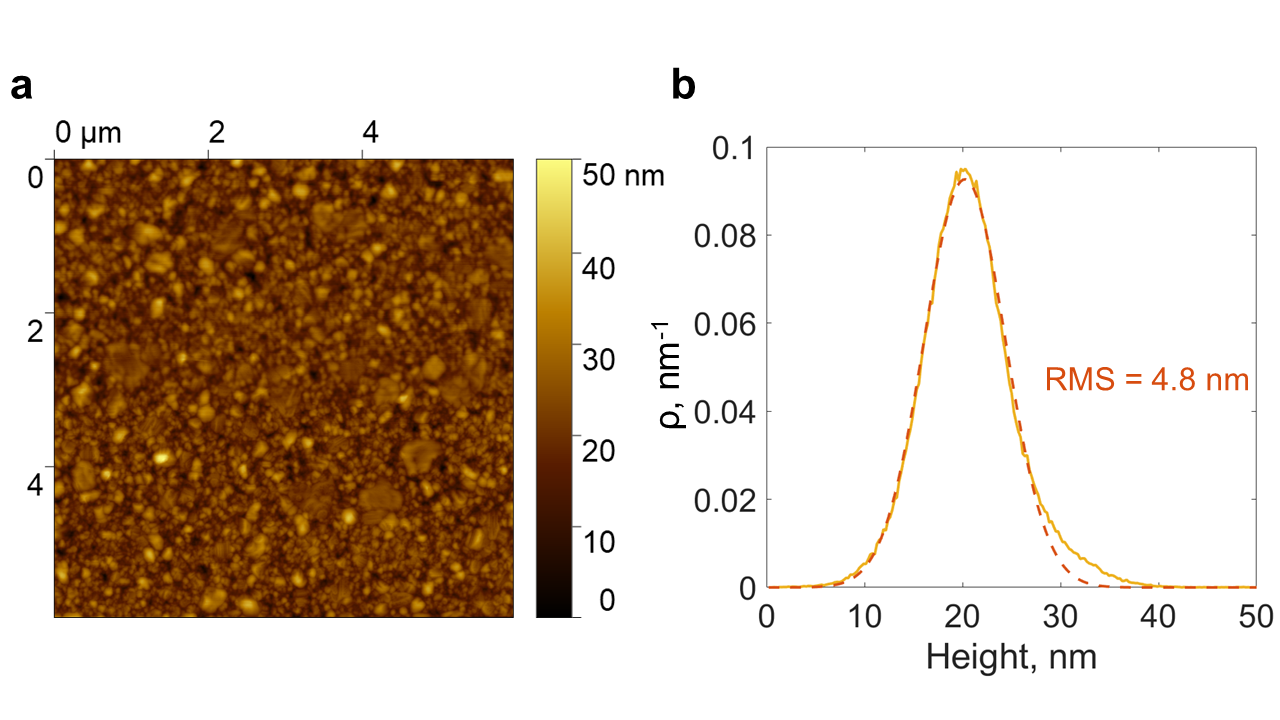}}
\caption{(a) AFM picture of the obtained metal-dielectric Al$_2$O$_3$/Ag/Si substrate morphology with width of thin Ag film of 50 nm (b) Hystogram of the film height extracted from the AFM measurement. Fitted Gaussian distribution give standart deviation equal to 4.8 nm.}
\label{figS_AFMAgfilm}
\end{figure}

\begin{figure}
\centering
\renewcommand\thefigure{S7}
\center{\includegraphics[width=0.8\linewidth]{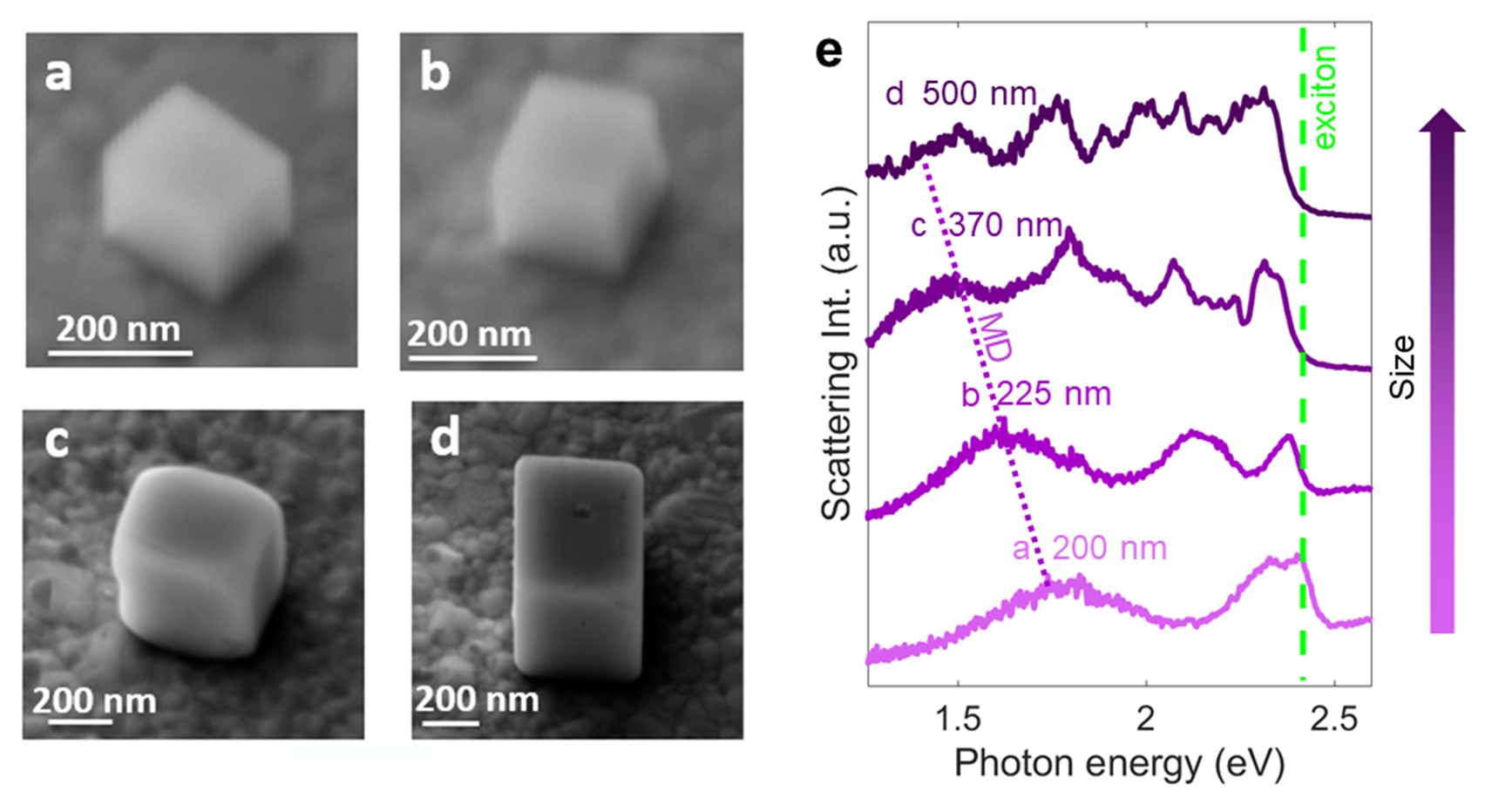}}
\caption{(a)-(d) SEM images of CsPbBr$_3$ cuboids with various characteristic sizes (a: 200 nm; b: 225 nm; c: 370 nm; d:500 nm) and (e) corresponding scattering spectra at room temperature. Green dashed line corresponds to exciton energy and purple dashed line to shift of magnetic dipole resonance.} 
\label{MieExp}
\end{figure}

Obtained nanocuboids suspension in hexane reveals bright green emission peaked at $\lambda_c$~524 nm with FWHM $\sim$ 21 nm upon UV lamp excitation, while extinction spectrum contains a sharp dip in the region of PL maximum (Fig\ref{Solut}a). This dip is attributed to Fano resonance~\cite{tiguntseva2018tunable}.
Notably, the suspension consists of nanocuboids of various hydrodynamic sizes from around 200 to 1000 nm which are slightly overestimated based on DLS measurements as compared to the real sizes (Fig\ref{Solut} b). According to the collected data the majority of nanocuboids have hydrodynamic diameter of about 296 nm. 
A suspension of perovskite cuboids was drop-casted on a metal-dielectric substrate (Fig.\ref{figS_AFMAgfilm}a-b). Regular shape and smooth facets of cavities with various dimensions were confirmed by SEM (Fig.\ref{MieExp}a-d). Scattering from single resonators was studied by means of dark-field spectroscopy at room temperature. Scattering spectra were measured by optical microscope (Carl Zeiss Axio Imager.A2m). The sample was illuminated in dark-field mode by a broadband halogen lamp (HAL 100) via 100x objective (EC Epiplan-NEOFLUAR). The scattered signal from single cuboid was collected by the same objective and sent to the spectrometer (Ocean Optics) through 100-$\mu$m optical fiber. The spectra showed sharp resonances of Mie-type which spectral position and order depends on cavity size. Therefore, small cuboids with size around 200 nm possesses pronounced magnetic dipole and quadrupole Mie-resonances. Increase in cavity size causes shift to low-energy of first-order resonances (purple dashed line) along with appearance of new high-order resonances near the exciton energy (green dashed line) (Fig.\ref{MieExp}e). Notably, the resultant scattering spectrum is a superposition of resonator’s Mie modes which allows to estimate the approximate geometrical parameters of cuboids. 

To solve the inverse scattering problem we applied a relevant numerical model based on the $T$-matrix method and the freely available software ``Smuthi''~\cite{Egel2014,Egel2016,Egel2017,Theobald2017,Egel2021}.
``Smuthi'' is a Python package for modeling electromagnetic scattering by one or more wavelength scale objects in layered systems efficiently and accurately.
The software integrates the $T$-matrix method of single-particle scattering with the scattering matrix formalism for electromagnetic field propagation through planar interfaces.
This method was used to find optimal multilayer systems designs and to obtain an extinction cross-section decomposition for all multipole components (in our case only for the dipole components).
To estimate the $T$-matrices of nonspherical particles, ``Smuthi'' uses the advanced NFM-DS (``Null-field Method with Discrete Sources'') FORTRAN~\cite{Doicu2006} software.
T-matrices computed with NFM-DS are processed in the ``Smuthi'' software to solve the problem of multiple scattering between such particles as well as between particles and layered media.

\begin{figure}
\centering
\renewcommand\thefigure{S8}
\includegraphics[width=6.4in]{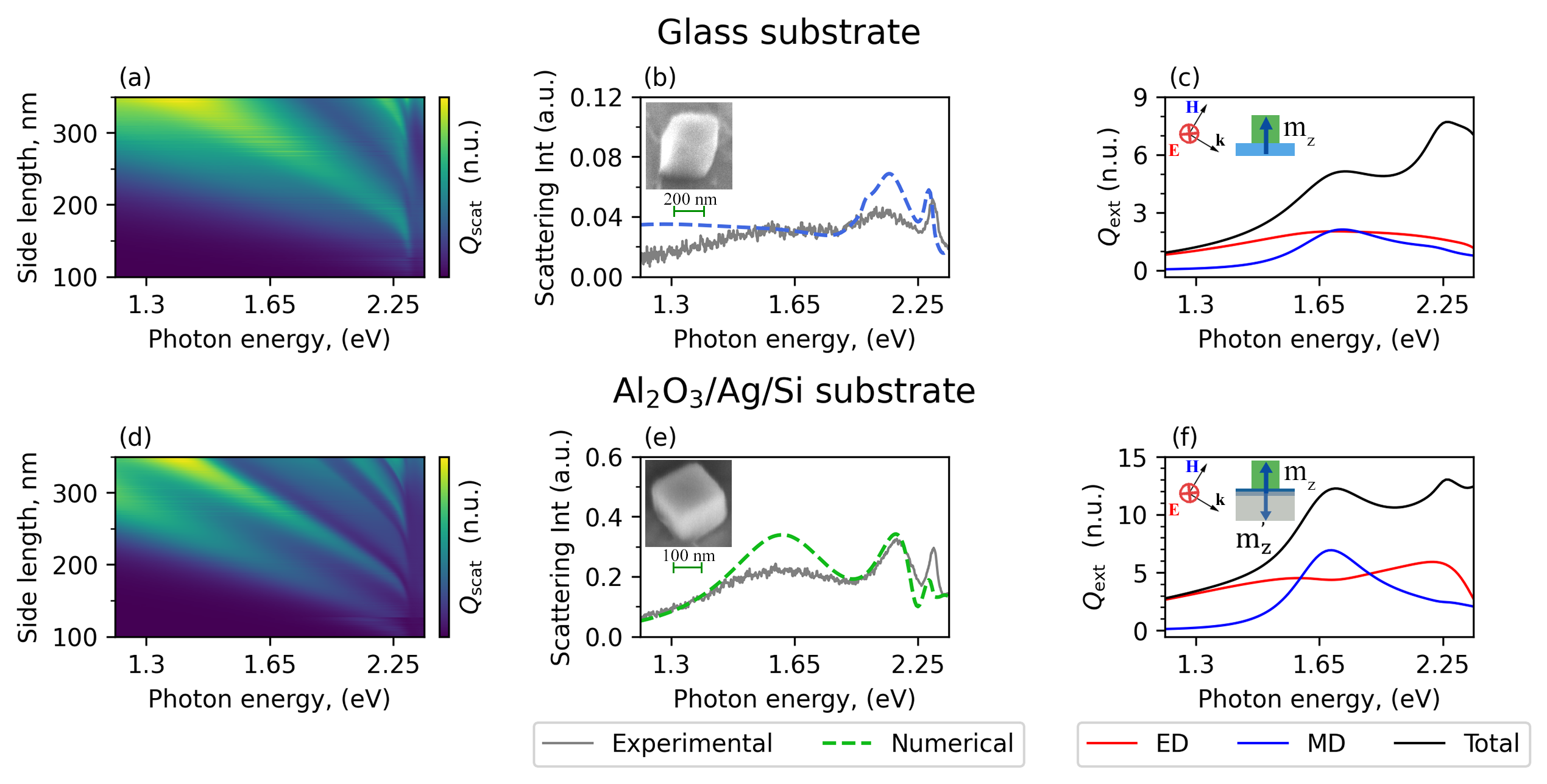}
\caption{Comparison of scattering properties of CsPbBr$_3$ cuboids on a glass (upper row) and on a metal-dielectric (bottom row) substrates. Scattering efficiency $Q_{\rm scat}$ for CsPbBr$_3$ nanocubes with different sizes on a glass substrate (a) and on a Al$_2$O$_3$/Ag/Si substrate (d).
Comparison of experimental (solid line) and numerical (dashed line) results of scattering spectra for the nanocuboid (320x370x260~nm) on a glass substrate (b) and the nanocuboid (205x225x195~nm) on a Al$_2$O$_3$/Ag/Si substrate (e).
Insets: SEM images of corresponding nanocuboids.
Electric (red curve) and magnetic (blue curve) dipole contributions in extinction efficiency $Q_{\rm ext}$ (black curve corresponds to total extinction) for 242~nm cube on a glass substrate (c) and on a Al$_2$O$_3$/Ag/Si substrate (f), respectively.
Insets: schematic representation of incident wave and formation of the image dipole moment.}
\label{fig:numerical_smth}
\end{figure}

As described earlier in the main text, metal-dielectric substrates are widely used to enhance field localization efficiency and mode selectivity~\cite{xifre2013mirror,sinev2016polarization}.
To demonstrate the promising potential of plasmonic substrates, we numerically simulate the scattering behavior of CsPbBr$_3$ nanocuboids on Al$_2$O$_3$/Ag/Si substrates (bottom row in Fig.~\ref{fig:numerical_smth}) and compare with similar nanocuboids on glass substrates (top row in Fig.~\ref{fig:numerical_smth}).

We evaluated a comparison of scattering spectra for a broad range of cube sizes (Fig.~\ref{fig:numerical_smth}a,d), revealing an order of magnitude higher scattering efficiency on the Al$_2$O$_3$/Ag/Si substrate, as well as a mode contrast. 
To validate the $T$-matrix method, the experimental scattering spectra obtained for similar CsPbBr$_3$ nanocuboids on the glass and Al$_2$O$_3$/Ag/Si substrates were compared with those predicted theoretically.
As shown in Fig.~\ref{fig:numerical_smth}b,e, obtained results demonstrate an excellent agreement between the numerical and experimental data, where the difference in amplitude for the resonance at high energy can be explained by the large absorption in this range. 
Notably, the scattering spectra were simulated for systems similar to the experimental ones, taking into account all input parameters of the experimental conditions including geometrical parameters (the difference in width, length and height).
Improved scattering efficiency and selectivity of modes on plasmonic substrates is related to the formation of mirror-image Mie modes~\cite{xifre2013mirror,sinev2016polarization}.
The multipole decomposition of extinction cross-section of a 242~nm CsPbBr$_3$ cube excited in TE mode is shown in (Fig.~\ref{fig:numerical_smth}c,f).
TE polarization of the incoming light leads to the efficient excitation of the $z$-oriented magnetic dipole and, in case of plasmonic substrate, formation of a magnetic dipole-image with anti-phase (inset in Fig.~\ref{fig:numerical_smth}f).
The interaction of anti-phase dipole sources leads to an increase in the intensity and $Q$-factor of magnetic dipole component in decomposition of the extinction cross-section for a cube on metal-dielectric substrate compared to the same cube on glass.  

Therefore, the integration of compact semiconductor resonators with metal-dielectric substrates is highly promising for lasing application. Moreover, the developed approach is highly useful for fast detection and primary characterization of nanocuboids, promising for lasing study, allowing to avoid premature SEM measurements, which destroys optical properties via increasing nonradiative traps.   

\section*{Section S7 Calculation of cold dispersion}
 Analytical dispersion of CsPbBr$_3$ at room and 6 K temperatures have been obtained from measurements of reflection coefficient of CsPbBr$_3$ microplates (MPs) by using of transfer matrix method (TMM).

\begin{figure}[h!]
\centering
\renewcommand\thefigure{S9}
\center{\includegraphics[width=0.8\linewidth]{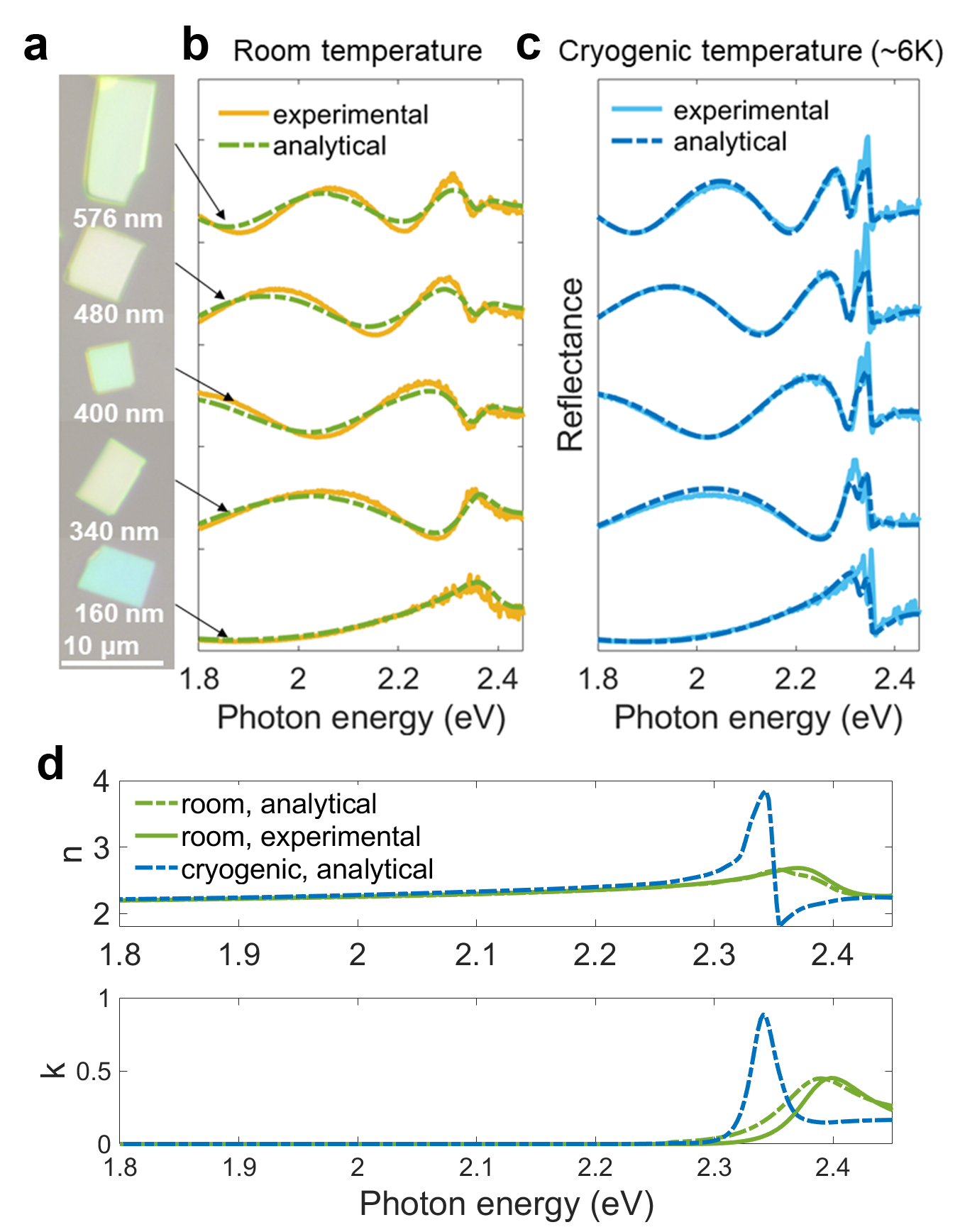}}
\caption{(a) Brightfield images of CsPbBr$_3$ plates with various thickness: 160 nm, 340 nm, 400 nm, 480 nm and 576 nm. Scale bar equals to 10 $\mu$m. (b) Experimental (continuous) and analytical (dashed)  reflection spectra of CsPbBr$_3$ plates with various thickness from (a) at room temperature. (c) Experimental (continuous) and analytical (dashed)  reflection spectra of CsPbBr$_3$ plates with various thickness from (a) at cryogenic ($\sim$6 K) temperature. (d) Real and imaginary part of CsPbBr$_3$ refractive index obtained from analytical calculation at room and cryogenic ($\sim$6 K) temperatures (dashed green and blue curve, respectively), whereas the experimental data at room temperature is taken from work \cite{ermolaev2022giant} (solid green curve).}
\label{coldnk}
\end{figure}

CsPbBr$_3$ MPs with various thicknesses (from 145 nm up to 1485 nm) were obtained on glass substrate according to the procedure reported by Zhizhchenko et al.~\cite{zhizhchenko2021direct}  (Fig.~\ref{coldnk}a). Reflection spectra were measured from 18 MPs at room and cryogenic temperatures upon broadband halogen lamp excitation at almost normal incidence. Reflected signal was collected from the center of MP through 100~$\mu$m-pinhole to avoid scattering from edges. The collected spectra were normalized to the incoming spectrum of source. Resulting experimental dependencies of reflection coefficient on photon energy obtained for selected five MPs with different thicknesses at room and cryogenic temperatures are shown in Fig.~\ref{coldnk}~b,c. 

To obtain the refractive index and the extinction coefficient of the perovskite MP on a glass substrate TMM has been employed. By treating the substrate as a semi-infinite space, the following analytical formula for the reflection coefficient can be derived:
\begin{equation}
    R^{\text{(TMM)}}_{t}(\lambda, T) = \left|
    \frac{
      \left(1 - \frac{n_{\text{plate}}(\lambda, T)}{n_0}\right)\left(1 + \frac{n_{\text{s}}}{n_{\text{plate}}(\lambda, T)}\right) e^{-i k_z t}
    + \left(1 + \frac{n_{\text{plate}}(\lambda, T)}{n_0}\right)\left(1 - \frac{n_{\text{s}}}{n_{\text{plate}}(\lambda, T)}\right) e^{i k_z t}
    }{
      \left(1 + \frac{n_{\text{plate}}(\lambda, T)}{n_0}\right)\left(1 + \frac{n_{\text{s}}}{n_{\text{plate}}(\lambda, T)}\right) e^{-i k_z t}
    + \left(1 - \frac{n_{\text{plate}}(\lambda, T)}{n_0}\right)\left(1 - \frac{n_{\text{s}}}{n_{\text{plate}}(\lambda, T)}\right) e^{i k_z t}
  }\right|^2,
\end{equation}
where $R^{\text{(TMM)}}(\lambda, T)$ is the intensity reflection coefficient depending on the wavelength and temperature, $t$ is the thickness of perovskite MP, $n_{\text{plate}}(\lambda, T)$ is the complex refractive index  of perovskite MP, $n_0$ is the refractive index of the surrounding atmosphere (we consider $n_0 = 1$), $n_{\text{s}}$ is the substrate refractive index (we consider $n_{\text{s}} = 1.55$), and 
\begin{equation}
    k_z = \frac{2\pi}{\lambda}\sqrt{n^2_{\text{film}}(\lambda, T) - n^2_0  \sin^2(\theta)},
\end{equation}
where $\theta$ is the incidence angle (we consider $\theta = 0$).

In order to parametrize the dispersion, we define the function $n_{\text{plate}}(\lambda)$ by its values at fixed wavelengths and use shape-preserving piecewise cubic interpolation to interconnect these points. Namely, we fix the refractive index at 20 uniformly distributed wavelengths in the 500 nm~$\leq\lambda<$~550~nm range, and the refractive index at 5 uniformly distributed points in the 550 nm~$\leq\lambda\leq$~750~nm range. The extinction coefficient is taken from direct measurements. Therefore we obtain a complex function $n_{\text{plate}}(\lambda, T)$ dependent on 24 fitting parameters (for a single temperature).

Additionally, due to the experimental errors, we introduce the following fitting parameters. First, due to the MP's surface corrugations, the measurement of the plate thicknesses with a profilometer has an experimental error of $\pm30~\text{nm}$, so we introduce the thickness correction $\Delta t_i$ for each plate, which is varied between $-30~\text{nm}$ and $+30~\text{nm}$. Second, due to the experimental errors in the reflection measurement, the reflection spectra were multiplied by the correction coefficients $K_i$ (for room temperature) and $K'_i$ (for 6~K), which became fitting parameters. The typical values of the correction coefficients lie between 0.9 and 1.1. The spectra are fitted together by minimizing the function $F$ by varying the total 102 fitting parameters:
\begin{equation}
    F = \sum_{i,j} \left|K_i R^{\text{(exp)}}_{t_i}(\lambda_j, 250~\text{K}) -  R^{\text{(TMM)}}_{t_i + \Delta t_i}(\lambda_j, 250~\text{K})\right|^2
    +
    \left|K'_i R^{\text{(exp)}}_{t_i}(\lambda_j, 6~\text{K}) -  R^{\text{(TMM)}}_{t_i + \Delta t_i}(\lambda_j, 6~\text{K})\right|^2,
\end{equation}
where $R^{\text{(exp)}}$ is the reflection coefficient measured experimentally.

The resulting reflection spectra are shown in Figs.~\ref{coldnk}~b,c labelled as `analytical'. The dispersions of MPs at room temperature and 6~K obtained by the fitting procedure described above are shown in Fig.~\ref{coldnk}~d. Notably, the dispersion obtained at room temperature is in a good agreement with one obtained experimentally for the same crystalline CsPbBr$_3$ material \cite{ermolaev2022giant}.

\begin{figure}
\centering
\renewcommand\thefigure{S10}
\center{\includegraphics[width=0.8\linewidth]{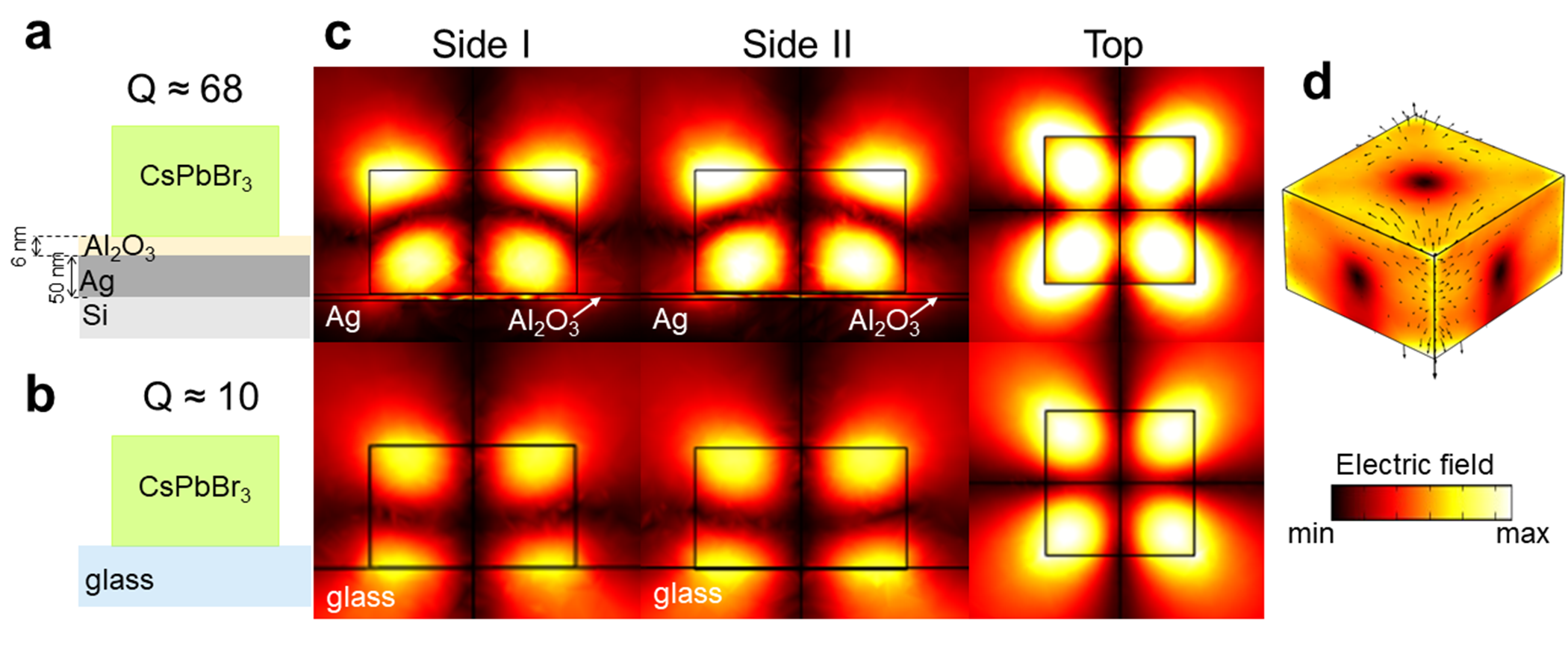}}
\caption{Effect of plasmonic substrate on EQ mode localization in 0.007$\mu m^3$ cuboid. Schematic images of CsPbBr$_3$ cuboid with the same size placed on metal-dielectric (a) and glass (b) substrates. (c) Normal electric field distribution in three different cross-sections (side I, side II, and top), where the upper row corresponds to cuboid on a plasmonic substrate (a), while the bottom one to glass (b). (d) Electric field distribution for the mode over the surface of the cuboid, where dominating contribution comes from EQ mode with azimuthal number m=2.}
\label{QDplandgl}
\end{figure}

\section*{Section S8. Photoluminescence decay measurements}
\begin{figure}[h!]
\centering
\renewcommand\thefigure{S11}
\center{\includegraphics[width=0.8\linewidth]{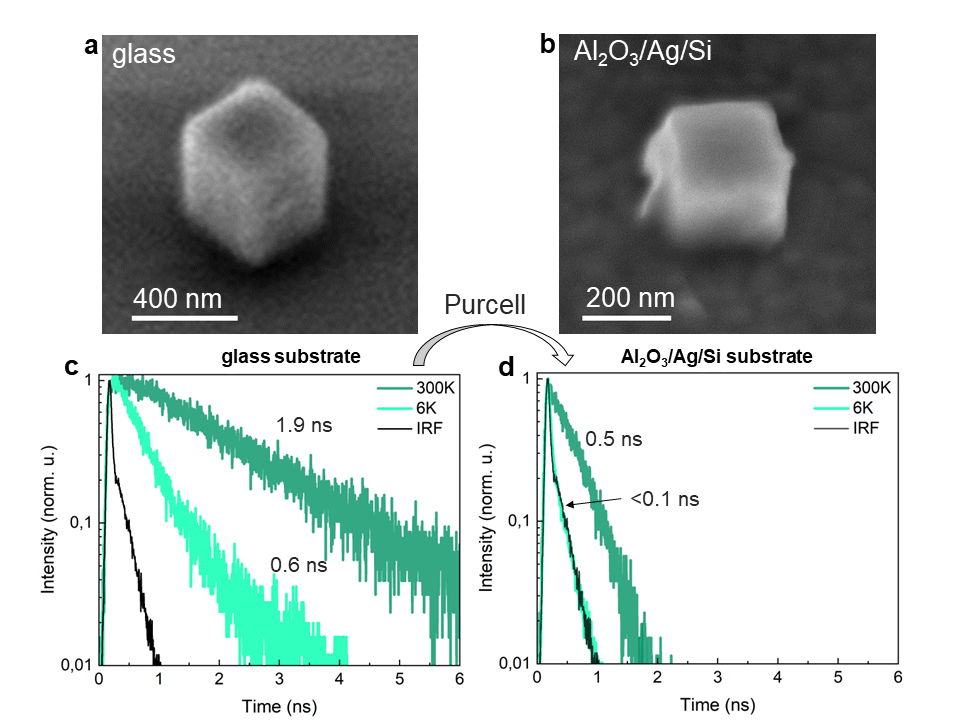}}
\caption{(a,b) SEM images of CsPbBr$_3$ nanocuboids on plasmonic and glass substrates, respectively. (c,d) Photoluminescence decay curves of CsPbBr$_3$ nanocuboids on plasmonic and glass substrates, respectively, obtained at room (300 K) and cryogenic (6 K) temperatures. Black decay corresponds to IRF.} 
\label{lifetime}
\end{figure}

To obtain lasing in perovskite cavities with physical volume less than 0.02 $\mu m^3$ they were integrated with metal-dielectric substrate which selectively enhances localization of Mie-resonances in nanocuboids due to a mirror-image effect \cite{xifre2013mirror},\cite{sinev2016polarization}. This enhancement related to Purcell effect and can be defined by Purcell factor $F_P$ proportional to $\frac{Q}{V_c}$ value and expressed as follows: $F_P = \frac{3}{4\pi^2}\frac{\lambda^3}{V_c}Q$, where $\lambda$ is a wavelength inside cavity, Q and  V$_c$  are quality factor and volume of mode, respectively. Experimentally, Purcell effect can be observed by shortening of the decay (lifetime) of spontaneous emission: $F_P=\gamma_R/\gamma$, where $\gamma_R$ and $\gamma$ – are radiative decay rates of dipole source in cavity or bulk, respectively. Therefore, PL decay in single cuboids with similar size on glass and plasmonic substrates (Figs.~\ref{lifetime} a-b) was studied in the same excitation conditions (pump fluence around 0.1 $\mu$uJ cm$^{-2}$, pulse duration around 220 fs, repetition rate equaled to 100 kHz and pump wavelength is centred to 490 nm) at room and cryogenic temperatures. At room temperature, PL lifetime equals to $\tau_{gl}$~=~1.9 ns and $\tau_{pl}$~=~0.5 ns for nanocuboids on glass and metal-dielectric substrates, respectively (Figs.~\ref{lifetime} c-d). At cryogenic temperature, lifetimes are getting nearly 3 times shorter for the glass substrate and at least more than 5 times shorter for plasmonic one that cannot be precisely resolved because of instrumental response (black curve in Figs.~S b-c) of the detector. Thus, the positive influence of metal-dielectric substrate on optical properties of perovskite nanocuboids was confirmed by time-resolved measurements.

\begin{table}[]
\begin{tabular}{|l|l|c|c|c|c|c|c|c|l|c|}
\hline
                              & \textbf{Type}                                                & \multicolumn{1}{c|}{\textbf{\begin{tabular}[c]{@{}l@{}}Active   \\ media\end{tabular}}} & \textbf{Cavity Size}                                                                 & \textbf{V ($\mu$m$^3$)}     & \textbf{V/ $\lambda^3$} & \textbf{Pump}                                                     & \textbf{Threshold}                                                & \textbf{Temp.} & \multicolumn{1}{c|}{\textbf{Ref}} & \textbf{Year} \\ \hline
\multirow{5}{*}{\textbf{UV}}  & Wire                                                         & \begin{tabular}[c]{@{}c@{}}GaN\\ 375 nm\end{tabular}                                                       & \begin{tabular}[c]{@{}c@{}}L:15 $\mu$m\\ D: 130 nm\end{tabular}              & 0.1                  & 2.08          & \begin{tabular}[c]{@{}c@{}}100 kHz\\ 10 ns\end{tabular}           & \begin{tabular}[c]{@{}c@{}}35 mJ/cm$^2$\end{tabular}  & RT             & \cite{zhang2014room}                                 & 2014           \\ \cline{2-11} 
                              & Wire                                                         & \begin{tabular}[c]{@{}c@{}}ZnO\\ 370 nm\end{tabular}                                                       & \begin{tabular}[c]{@{}c@{}}L: 1.24 $\mu$m\\ H=W: 28 nm\end{tabular}                & 0.003                & 0.05          & \begin{tabular}[c]{@{}c@{}}1 kHz\\ 0.5 ns\end{tabular}            & \begin{tabular}[c]{@{}c@{}}30 mJ/cm$^2$\end{tabular}    & 77 K            & \cite{chou2015ultrastrong}                                 & 2015          \\ \cline{2-11} 
                              & Wire                                                         & \begin{tabular}[c]{@{}c@{}}ZnO\\380 nm\end{tabular}                                                       & \begin{tabular}[c]{@{}c@{}}L: 1.7 $\mu$m\\ H=W: 30 nm\end{tabular}                 & 0.004                & 0.072         & \begin{tabular}[c]{@{}c@{}}1 kHz\\ 0.5 ns\end{tabular}            & \begin{tabular}[c]{@{}c@{}}55 mJ/cm$^2$\end{tabular}  & RT             & \cite{chou2016high}                                 & 2016          \\ \cline{2-11} 
                              & Wire                                                         & \begin{tabular}[c]{@{}c@{}}ZnO\\ 370 nm\end{tabular}                                                       & \begin{tabular}[c]{@{}c@{}}L: 1.5 $\mu$m\\ H=W: 30 nm\end{tabular}                 & 0.004                & 0.069         & \begin{tabular}[c]{@{}c@{}}1 kHz \\ 0.5 ns\end{tabular}            & \begin{tabular}[c]{@{}c@{}}27 mJ/cm$^2$ \end{tabular} & 77 K           & \cite{chou2018ultracompact}                                 & 2018          \\ \cline{2-11} 
                              & Wire                                                         & \begin{tabular}[c]{@{}c@{}}ZnO\\ 372 nm\end{tabular}                                                       & \begin{tabular}[c]{@{}c@{}}L: 0.9 $\mu$m\\ H=W: 35 nm\end{tabular}                 & 0.003                & 0.06          & \begin{tabular}[c]{@{}c@{}}1 kHz \\ 0.5 ns\end{tabular}            & \begin{tabular}[c]{@{}c@{}}45 mJ/cm$^2$\end{tabular}       & 77 K           & \cite{cheng2018high}                                 & 2018          \\ \hline
\multirow{7}{*}{\textbf{VIS}} & 
\begin{tabular}[c]{@{}l@{}}Bundle\\of rods\end{tabular}     & 
\begin{tabular}[c]{@{}c@{}}InGaN/GaN\\ 533 nm\end{tabular}                                                 & \begin{tabular}[c]{@{}c@{}}L: 680 nm\\ H=W: 530 nm\end{tabular}     & 0.13                 & 0.44          & \begin{tabular}[c]{@{}c@{}}90 MHz \\ 150 fs\end{tabular}          & \begin{tabular}[c]{@{}c@{}}3   mJ/cm$^2$\end{tabular} & 7 K            & \cite{wu2011plasmonic}                                 & 2011          \\ \cline{2-11}                      &
\begin{tabular}[c]{@{}l@{}}Wire\end{tabular}       & \begin{tabular}[c]{@{}c@{}}MAPbI$_3$\\ 780 nm\end{tabular}                                                      & \begin{tabular}[c]{@{}c@{}}L: 7.5 $\mu$m \\ H: 600 nm \\W: 300 nm \end{tabular}                & 1.35                 & 2.84           & \begin{tabular}[c]{@{}c@{}}250 kHz \\ 150 fs\end{tabular}           & \begin{tabular}[c]{@{}c@{}}0.4 $\mu$J/cm$^2$\end{tabular}   & RT             & \cite{zhu2015lead}            & 2015        \\ \cline{2-11} 
                              & \begin{tabular}[c]{@{}c@{}}Square\\plate\end{tabular}                                                         & \begin{tabular}[c]{@{}c@{}}MAPbBr$_3$\\ 557 nm\end{tabular}                                                       & \begin{tabular}[c]{@{}c@{}}L=W: 2 $\mu$m\\ H: 0.6 $\mu$m\end{tabular}                & 2.4                & 13.89          & \begin{tabular}[c]{@{}c@{}}1 kHz\\ 120 fs\end{tabular}            & \begin{tabular}[c]{@{}c@{}}3.6 $\mu$J/cm$^2$\end{tabular}    & RT            & \cite{liao2015perovskite}                                 & 2015        \\ \cline{2-11}                      &
\begin{tabular}[c]{@{}l@{}}Wire\end{tabular}       & \begin{tabular}[c]{@{}c@{}}MAPbI$_3$\\ 768 nm\end{tabular}                                                      & \begin{tabular}[c]{@{}c@{}}L: 5.1 $\mu$m \\ H: 126 W: 167 nm \end{tabular}                & 0.1                 & 0.24           & \begin{tabular}[c]{@{}c@{}}1 kHz \\ 120 fs\end{tabular}           & \begin{tabular}[c]{@{}c@{}}29 $\mu$J/cm$^2$\end{tabular}   & RT             & \cite{yu2016organic}            & 2016         \\ \cline{2-11} 
                              & \begin{tabular}[c]{@{}l@{}}Sphere\end{tabular}    & \begin{tabular}[c]{@{}c@{}}CsPbBr$_3$\\ 545.2 nm\end{tabular}                                                    & \begin{tabular}[c]{@{}c@{}} D: 780 nm\end{tabular}                & 0.48                   & 2.95          & \begin{tabular}[c]{@{}c@{}}10 kHz \\ 40 fs\end{tabular}             & 0.42 $\mu$J/cm$^2$                                                       & RT             & \cite{tang2017single}            & 2017 \\ \cline{2-11}                      &
\begin{tabular}[c]{@{}l@{}}Square\\ plate\end{tabular}       & \begin{tabular}[c]{@{}c@{}}CdSe\\ 700 nm\end{tabular}                                                      & \begin{tabular}[c]{@{}c@{}}H: 137 nm\\ L=W: 1 $\mu$m\end{tabular}                & 0.14                 & 0.4           & \begin{tabular}[c]{@{}c@{}}1 kHz \\ 4.5 ns\end{tabular}           & \begin{tabular}[c]{@{}c@{}}45 mJ/cm$^2$\end{tabular}   & RT             & \cite{wang2017unusual}            & 2017        
                            \\ \cline{2-11} 
                              & \begin{tabular}[c]{@{}l@{}}Hex.\\ plate\end{tabular}    & \begin{tabular}[c]{@{}c@{}}MAPbI$_3$\\ 770 nm\end{tabular}                                                    & \begin{tabular}[c]{@{}c@{}}H: 50 nm\\ L: 10 $\mu$m\end{tabular}                & 13                   & 28.5          & \begin{tabular}[c]{@{}c@{}}1 kHz \\ 100 fs\end{tabular}             & 59.2 $\mu$J/cm$^2$                                                       & RT             & \cite{huang2018formation}            & 2018          
                                  \\ \cline{2-11} 
                              & Cuboid                                                         &  \begin{tabular}[c]{@{}c@{}}CsPbBr$_3$\\ 540 nm\end{tabular}                                                   & W=L=H: 400 nm                                                                       & 0.064                & 0.49          & \begin{tabular}[c]{@{}c@{}}1 kHz\\ 35 fs\end{tabular}             & 40.2 $\mu$J/cm$^2$                                  & RT             & \cite{liu2018robust}                                 & 2018      \\ \cline{2-11} & \begin{tabular}[c]{@{}l@{}}Wire\end{tabular}    & \begin{tabular}[c]{@{}c@{}}CsPbBr$_3$\\ 524 nm\end{tabular}                                                    & \begin{tabular}[c]{@{}c@{}}W=H: 250 nm\\ L: 2 $\mu$m\end{tabular}                & 0.125                   & 0.9          & \begin{tabular}[c]{@{}c@{}}CW\end{tabular}             & 46 mW/cm$^2$                                                       & 4K             & \cite{jiang2018continuous}            & 2018    
                                 \\ \cline{2-11} & \begin{tabular}[c]{@{}l@{}}Wire\end{tabular}    & \begin{tabular}[c]{@{}c@{}}CsPbBr$_3$\\ 534 nm\end{tabular}                                                    & \begin{tabular}[c]{@{}c@{}}L: 3.5 $\mu$m\\ H: 120 nm\\ W: 200 nm\end{tabular}                & 0.084                   & 0.55          & \begin{tabular}[c]{@{}c@{}}1 kHz \\ 100 fs\end{tabular}             & 12 $\mu$J/cm$^2$                                                       & RT             & \cite{wu2018all}            & 2018    
                                 \\ \cline{2-11} 
                              & Cuboid                                                        & \begin{tabular}[c]{@{}c@{}}CsPbBr$_3$\\ 533 nm\end{tabular}                                                   & W=L=H: 310 nm                                                                       & 0.03                 & 0.197         & \begin{tabular}[c]{@{}c@{}}100 kHz\\ 150fs\end{tabular}           & 300 $\mu$J/cm$^2$                                    & RT             & \cite{tiguntseva2020room}                                & 2020          \\ \cline{2-11} 
                              & Cuboid                                                         & \begin{tabular}[c]{@{}c@{}}CsPbBr$_3$\\ 540 nm\end{tabular}                                                   & \begin{tabular}[c]{@{}c@{}}W: 580 nm\\  L: 560 nm \\H: 320 nm  \end{tabular}                                                                 & 0.1                  & 0.66          & \begin{tabular}[c]{@{}c@{}}20 kHz\\ 4 ns\end{tabular}              & 0.48 mJ/cm$^2$                                   & RT             & \cite{cho2021submicrometer}                                & 2021        \\ \cline{2-11} 
                              & Wire                                                         & \begin{tabular}[c]{@{}c@{}}CsPbBr$_3$\\ 534 nm\end{tabular}                                                   &
                              \begin{tabular}[c]{@{}c@{}}L: 4.5 $\mu$m\\ W: 260 nm\\ H: 290 nm    \end{tabular} 
                                                                                                & 0.34                  & 2.23          & \begin{tabular}[c]{@{}c@{}}50 kHz\\ 150 fs\end{tabular}              & 24 $\mu$J/cm$^2$                                   & RT             & \cite{safronov2022efficient}                                & 2022          \\ \cline{2-11} 
                              & \textbf{Cuboid}                                                & \textbf{\begin{tabular}[c]{@{}c@{}}CsPbBr$_3$\\ 533 nm\end{tabular}}                                          & \textbf{\begin{tabular}[c]{@{}c@{}}W: 215 nm\\ L: 225 nm\\ H: 140 nm\end{tabular}
                              }                                                          & \textbf{0.007} & \textbf{0.04} & \textbf{\begin{tabular}[c]{@{}c@{}}100 kHz\\ 220 fs\end{tabular}} & \textbf{1 $\mu$J/cm$^2$}                               & \textbf{6.4 K} & \textbf{*}                       & \textbf{2023} \\ \hline
\multirow{2}{*}{\textbf{NIR}} & \begin{tabular}[c]{@{}l@{}}Plasm.\\ disk\end{tabular}     & \begin{tabular}[c]{@{}c@{}}InGaAsP\\ 1430 nm\end{tabular}                                                  & \begin{tabular}[c]{@{}c@{}}D: 920 nm\\ H: 1.28 $\mu$m\end{tabular} & 0.2                  & 0.07          & \begin{tabular}[c]{@{}c@{}}300 kHz\\ 12 ns\end{tabular}             & \begin{tabular}[c]{@{}c@{}}1 mJ/cm$^2$\end{tabular}    & RT             & \cite{nezhad2010room}                                & 2010          \\ \cline{2-11} 
                              & \begin{tabular}[c]{@{}l@{}}Plasm.\\ cylinder\end{tabular} & \begin{tabular}[c]{@{}c@{}}InGaAsP\\ 1400 nm\end{tabular}                                                  & \begin{tabular}[c]{@{}c@{}}D: 400 nm \\H: 210 nm\end{tabular} & 0.03                 & 0.01          & CW                                                                & Thresholdless                                                     & 4.5 K          & \cite{khajavikhan2012thresholdless}                                & 2012      
                              \\ \cline{2-11} 
                              & \begin{tabular}[c]{@{}l@{}}Cylinder\end{tabular} & \begin{tabular}[c]{@{}c@{}}GaAs\\ 825 nm\end{tabular}                                                  & \begin{tabular}[c]{@{}c@{}}D: 500 nm \\H: 330 nm\end{tabular} & 0.07                 & 0.1          & \begin{tabular}[c]{@{}c@{}}20 kHz\\ 200 fs\end{tabular}                                                                & 260 $\mu$J/cm$^2$
                                                     & 77 K          & \cite{mylnikov2020lasing}                                & 2020    
                                                       \\ \cline{2-11} 
                              & \begin{tabular}[c]{@{}l@{}}Disk\end{tabular} & \begin{tabular}[c]{@{}c@{}}WS$_2$\\ 868.5 nm\end{tabular}                                                  & \begin{tabular}[c]{@{}c@{}}D: 3 $\mu$m \\H: 50 nm\end{tabular} & 0.35                 & 0.58         & CW                                                                & 
                                   1.25 kW/cm$^2$
                  & RT          & \cite{sung2022room}                                & 2022    
                              \\ \hline
\end{tabular}
\caption{Comparison table of nanolasers for the last 12 years. This work is highlighted by the bold font. In the table L - length, D - diameter, H - height, W - width, * - this work. }
\label{table:1}
\end{table}

\begin{figure}
\centering
\renewcommand\thefigure{S12}
\center{\includegraphics[width=0.8\linewidth]{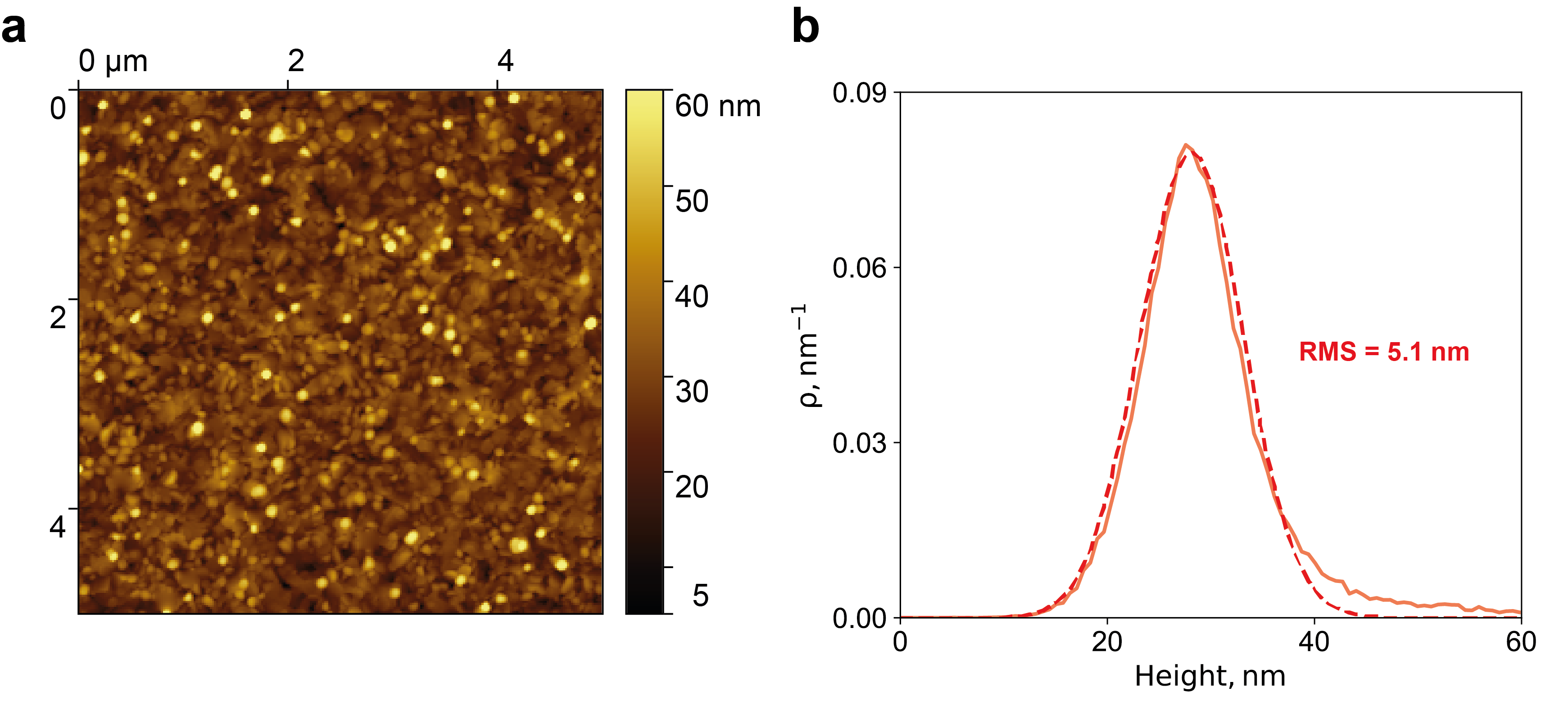}}
\caption{(a) AFM picture of the studied lead-bromide perovskite thin film morphology with width of 120 nm (b) Hystogram of the film height extracted from the AFM measurement. Fitted Gaussian distribution give standart deviation equal to 5.1 nm.}
\label{figS_AFMfilm}
\end{figure}

\begin{figure}
\centering
\renewcommand\thefigure{S13}
\center{\includegraphics[width=0.7\linewidth]{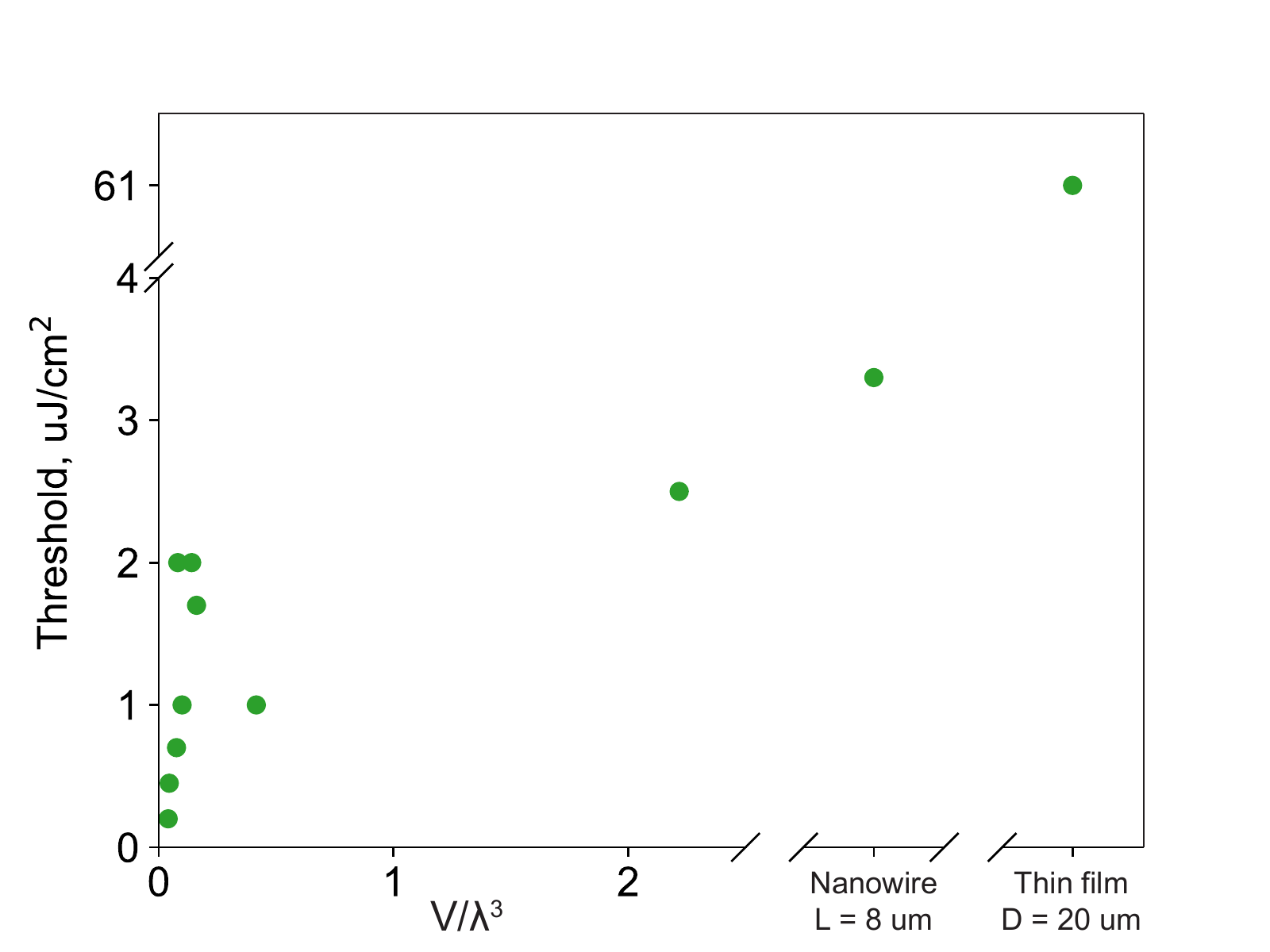}}
\caption{Stimulated emission pump fluence threshold of studied perovskite structures at 6 K depending on the size. For nanocubes size normalized on emission wavelength $V/\lambda^3$ is shown. Threshold of multilasing in nanowire was determined, L is the length of studied nanowire with lateral cross section around 600 nm. Threshold of studied perovsite thin film with width of 100 nm was determined, diameter of pumping spot D is equal to 20 $\mu$m.}
\label{fig_thre_size}
\end{figure}

\begin{figure}
\centering
\renewcommand\thefigure{S14}
\center{\includegraphics[width=1\linewidth]{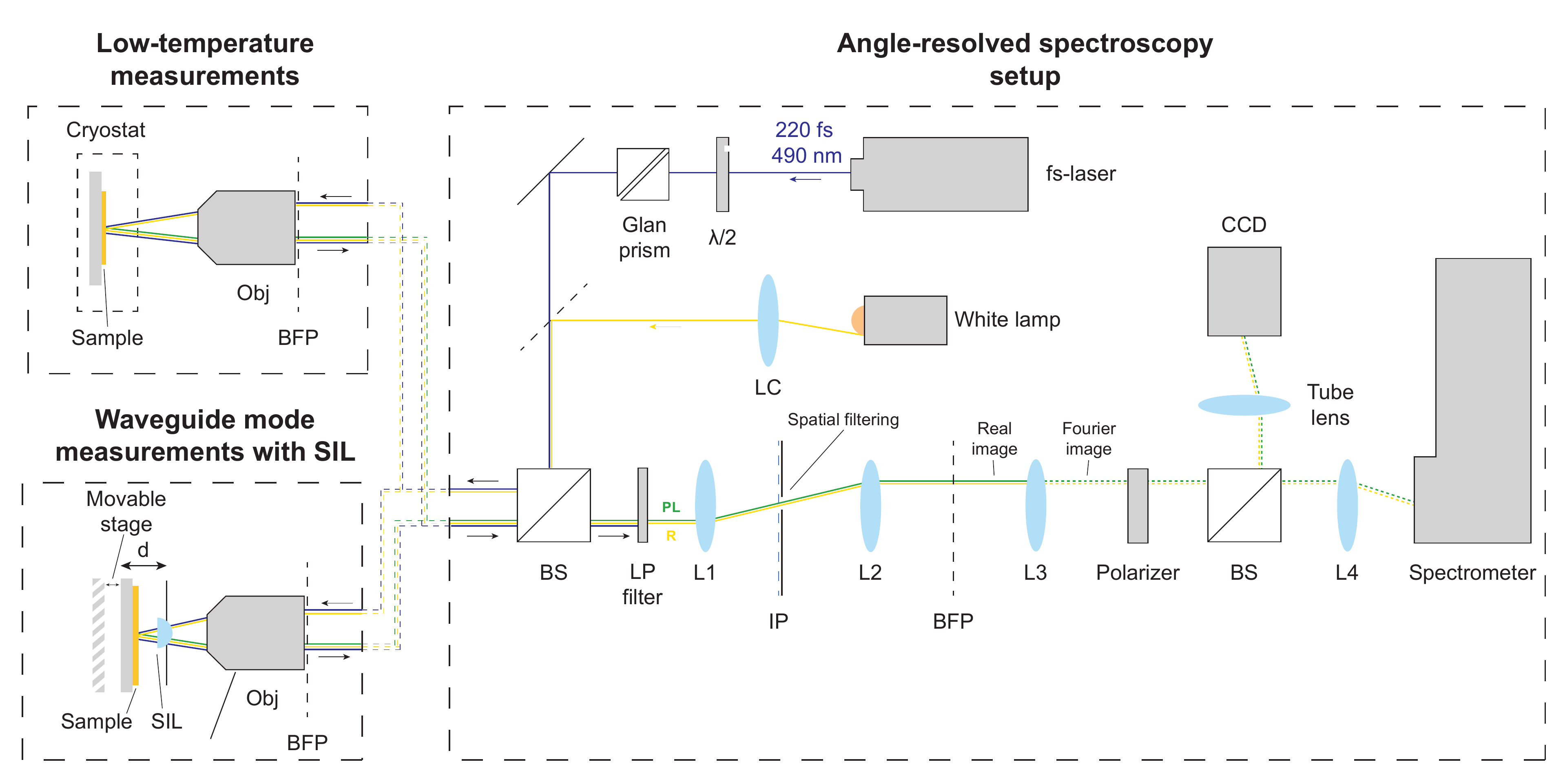}}
\caption{Scheme of the optical experimental setup.}
\label{figS_expSetup}
\end{figure}

\end{document}